\documentclass[english,aps,amsfonts,amsmath,prd,preprint,nofootinbib]{revtex4}
\pdfoutput=1
\usepackage[T1]{fontenc}
\usepackage[latin9]{inputenc}
\usepackage{amsmath}
\usepackage{amssymb}
\usepackage{graphicx}
\usepackage{esint}

\makeatletter

\providecommand{\tabularnewline}{\\}

\@ifundefined{textcolor}{}
{%
 \definecolor{BLACK}{gray}{0}
 \definecolor{WHITE}{gray}{1}
 \definecolor{RED}{rgb}{1,0,0}
 \definecolor{GREEN}{rgb}{0,1,0}
 \definecolor{BLUE}{rgb}{0,0,1}
 \definecolor{CYAN}{cmyk}{1,0,0,0}
 \definecolor{MAGENTA}{cmyk}{0,1,0,0}
 \definecolor{YELLOW}{cmyk}{0,0,1,0}
}

\usepackage{babel}

\usepackage{babel}

\usepackage{babel}

\usepackage{babel}

\usepackage{babel}

\usepackage{babel}

\makeatother

\usepackage{babel}
\begin{document}

\title{Primordial black hole and wormhole formation by domain walls}

\author{Heling Deng$^{a}$, Jaume Garriga$^{a,b}$ and Alexander Vilenkin$^{a}$}

\affiliation{$^{a}$ Institute of Cosmology, Tufts University, 574 Boston Ave,
Medford, MA, 02155 U.S.A.}

\affiliation{$^{b}$ Departament de Fisica Fonamental i Institut de Ciencies del
Cosmos, Universitat de Barcelona, Marti i Franques, 1, Barcelona,
08028 Spain }
\begin{abstract}
In theories with a broken discrete symmetry, Hubble sized spherical
domain walls may spontaneously nucleate during inflation. These objects
are subsequently stretched by the inflationary expansion, resulting
in a broad distribution of sizes. The fate of the walls after inflation
depends on their radius. Walls smaller than a critical radius fall
within the cosmological horizon early on 
and collapse due to their own tension, forming ordinary black holes.
But if a wall is large enough, its repulsive gravitational field becomes
dominant much before the wall can fall within the cosmological horizon.
In this ``supercritical'' case, a wormhole throat develops, connecting
the ambient exterior FRW universe with an interior baby universe,
where the exponential growth of the wall radius takes place. The wormhole
pinches off in a time-scale comparable to its light-crossing time,
and black holes are formed at its two mouths. As discussed in previous
work, the resulting black hole population has a wide distribution
of masses and can have significant astrophysical effects. The mechanism
of black hole formation has been previously studied for a dust-dominated
universe. Here we investigate the case of a radiation-dominated universe,
which is more relevant cosmologically, by using numerical simulations
in order to find the initial mass of a black hole as a function of
the wall size at the end of inflation. For large supercritical domain
walls, this mass nearly saturates the upper bound according to which
the black hole cannot be larger than the cosmological horizon. We
also find that the subsequent accretion of radiation satisfies a scaling relation, resulting in a mass increase by about a factor of 2. 
\end{abstract}
\maketitle

\section{Introduction}

Quantum fluctuations of matter fields during the inflationary epoch
may have seeded the large scale structure we observe in the universe.
While this process is usually considered in the perturbative regime,
it was recently pointed out in Ref. \cite{Garriga} that non-perturbative
quantum effects could also contribute to the formation of structure
on astrophysically relevant scales. Here we will focus on the scenario
where the spontaneous nucleation of spherical domain walls during
inflation leads to a population of primordial black holes, with a
wide spectrum of masses.

Domain walls are two dimensional topological defects that may arise
in theories with a spontaneously broken discrete symmetry, separating
neighboring regions where the field is in different vacua. They are
characterized by a finite energy density $\sigma$ per unit surface,
and a tension of equal magnitude. Such tension, or negative pressure,
is the source of a repulsive gravitational field. In the thin wall
limit, the metric for a planar domain wall in the $yz$-plane (in
an otherwise empty space) is given by \cite{Vilenkin,Alex,Ipser}
\begin{equation}
ds^{2}=-\left(1-\frac{\left|x\right|}{t_{\sigma}}\right)^{2}dt^{2}+dx^{2}+\left(1-\frac{\left|x\right|}{t_{\sigma}}\right)^{2}e^{2t/t_{\sigma}}\left(dy^{2}+dz^{2}\right),
\end{equation}
where 
\begin{equation}
t_{\sigma}=\frac{1}{2\pi G\sigma}.\label{tsigma}
\end{equation}
Here, $G$ is Newton's constant. The $(x,t)$-part of the metric is
a $(1+1)$-dimensional Rindler space, and a geodesic observer near
$x=0$ will recede from the wall with acceleration $t_{\sigma}^{-1}$.
Also, the hypersurface $x=0$, corresponding to the worldsheet of
the wall, inflates at the constant rate $H_{\sigma}=t_{\sigma}^{-1}$.

Cosmologically, domain walls may form by the Kibble mechanism at a
phase transition, when the universe goes from a high temperature symmetric
phase to a broken phase at lower temperatures. In this case, the distribution
of defects at the time of formation will include closed walls as well
as infinite walls of quite irregular shapes \cite{Vilenkin}. After
the phase transition, the number of walls stretching across a horizon
region is at least of order one, and such defects would come to dominate
over the radiation or matter density at a time of order $t_{\sigma}$.
This leads to a domain wall problem unless $t_{\sigma}\gtrsim t_{0}$
where $t_{0}$ is the current age of the universe. This severely limits
the tension to be rather low $\sigma\lesssim1/Gt_{0}\sim(100\ {\rm MeV})^{3}$,
while an even stronger constraint, $\sigma\lesssim(1\ {\rm MeV})^{3}$
follows from the observed isotropy of the CMB \cite{paul}.

Here we will be interested in a very different production mechanism,
where the domain wall problem does not arise, and the above constraints
on $\sigma$ do not apply. The scenario of our interest is the spontaneous
nucleation of domain walls during inflation \cite{Basu}. This can
be thought of as quantum process by which the radius of a closed wall
tunnels from vanishing size to Hubble size. The walls thus created
are subsequently stretched by the inflationary expansion, resulting
in a broad distribution of sizes. In the semiclassical limit, they
are expected to be nearly spherical, with a size distribution which
at the end of inflation takes the form, 
\begin{equation}
dn\approx\lambda(R)\frac{dR}{R^{4}}.\label{dist}
\end{equation}
Here, $n$ is the number density of walls, and $\lambda$ is the nucleation
rate in Hubble units during inflation. The nucleation rate has an
exponential dependence on the action $S_{E}$ of the semiclassical
tunneling path, $\lambda\propto e^{-S_{E}}$, where $S_{E}\approx2\pi^{2}\sigma H_{i}^{-3}$.
Here $H_{i}$ is the inflationary Hubble rate. The radii $R$ are
in the range $1\lesssim H_{i}R\lesssim e^{N}$ , where $N\sim60$
is the number of e-folding of slow roll inflation within observationally
accessible scales. For constant $\lambda$ the distribution (\ref{dist})
would be scale invariant, but we may expect some dependence of $\lambda$
on $R$, since the value of $H_{i}$, and perhaps also $\sigma$,
will slowly vary during inflation.

The dynamics of spherical domain walls in vacuum has been thoroughly
studied in the literature. Ref. \cite{Ipser} considered the evolution
of a spherical domain wall in asymptotically flat space. In this case
the interior metric is Minkowski and the exterior is Schwarzschild.
It was shown that if the wall is initially larger than the Schwarzschild
radius, then it always collapses to a black hole singularity. Refs.
\cite{Sato,BKT,Blau} considered the more general case of a false
vacuum bubble surrounded by true vacuum, and studied the motion of
the domain wall at the boundary of these two regions by matching an
interior de Sitter with a Schwarzschild exterior metric. This includes,
in particular, the case when the energy density inside the bubble
vanishes, so that the only source of the gravitational field is the
domain wall. A solution was found where a domain wall emerges from
an initial ``white hole'' singularity, and expands without bound
thereafter in a baby universe. The baby universe is initially connected
to the asymptotically flat region by a wormhole, but the wormhole
eventually pinches off, leading to a change in topology. Because of
the initial white hole singularity, such solutions may be considered
somewhat unphysical. However, as pointed out in Ref. \cite{Garriga},
a similar process can occur for sufficiently large walls in the aftermath
of a standard inflationary phase\footnote{The possibility of wormhole formation in cosmological spacetimes has
also been discussed earlier in Refs. \cite{Harada,Carr1,Carr2}, but
without suggesting a realistic cosmological scenario where it can
be realized.}. In this case there is no initial white hole singularity.

In our cosmological setting, the fate of the walls depends on their
size $R$ at the time $t_{i}$ when inflation ends \cite{Garriga}.  At that time, all walls have radii greater than the horizon and are essentially at rest with respect to the Hubble flow.
A useful parameter is the time $t_{H}(r_{i})$ when the comoving scale
$r_{i}=R/a(t_{i})$ would fall within the Hubble radius in an undisturbed
FRW universe. In a dust or radiation-dominated universe, we have $t_{H}={8r_{i}^{3}}/({27t_{i}^{2}})$,
or $t_{H}={r_{i}^{2}}/({4t_{i}})$ respectively. Here, and for the
rest of this paper, we will adopt the convention that $a(t_{i})=1$,
so that $r_{i}=R$.

A wall with $t_{H}(r_{i})\ll t_{\sigma}$ is called ``subcritical'',
and its gravitational field can be safely neglected before horizon
crossing. The wall grows for a while due to the cosmological expansion,
and at the time $t_{H}$ it realizes that it's actually a ball. Then
it is forced by the surface tension to shrink to a size smaller than
the corresponding Schwarzschild radius, forming an ordinary black
hole. Afterwards, some of the nearby fluid falls into the black hole,
leading to mass accretion.

On the other hand, a wall with $t_{H}(r_{i})\gg t_{\sigma}$ is called
``supercritical'', and its gravitational effect becomes significant
at time $t_{\sigma}$, before the wall has a chance to fall within
the cosmological horizon. Due to its repulsive nature, the wall pushes
fluid away, leaving two nearly empty layers in its vicinity, which
separate the wall from an interior and an exterior FRW regions. The
exterior FRW universe continues its power law Hubble expansion, but
the wall radius grows exponentially in proper time. This is only possible
because the wall creates a wormhole, through which it escapes into
a baby universe. The wormhole pinches off in a timescale comparable
to its light crossing time, and observers on either side of the wormhole
throat will see a black hole forming, possibly with different masses
on both sides.

The reason why there is no domain wall problem in this scenario can
easily be understood as follows. Subcritical walls form black holes
before their contribution to the energy (on the Hubble scale) can
become significant. After that, they simply behave as a contribution
to cold dark matter. On the other hand, for supercritical walls, the
repulsive gravitational field becomes important at the time $t_{\sigma}$,
when their energy density locally dominates over the matter density
(in all Hubble patches containing a segment of the wall). As soon
as this happens, the wall starts creating a wormhole, and the ``domain
wall problem'' is exported into the resulting baby universe. All
we are left with in the parent universe is a distribution of primordial
black holes with a broad spectrum of masses \cite{Garriga}.

The relative contribution of black holes of mass $M_{bh}$ to the
dark matter density $\rho_{m}$ is given by $f\equiv(M_{bh}/\rho_{m})dn(M_{bh})/d\log M_{bh}$.
This is determined by the initial distribution of wall sizes, Eq.
(\ref{dist}), by using the relation $M_{bh}(R)$ which gives the
black hole mass a function of the initial wall radius $R$. In Ref.
\cite{Tanahashi} this relation was studied numerically for subcritical
walls in a dust-dominated background. In Ref. \cite{Garriga} the
limits of small and large walls were studied analytically for the
case of a dust background. The limit of small walls in a radiation
background was also discussed. Nonetheless, the more interesting case
of large supercritical walls in a radiation-dominated universe was
not analyzed quantitatively. In this case, pressure gradients create
a flux of radiation across the wormhole, which affects the mass of
the resulting black hole, but the impact of this effect is hard to
estimate analytically.

In Ref. \cite{Garriga}, the mass of supercritical black holes was
tentatively assumed to be of order $M_{bh}\sim t_{H}/G$, saturating
the condition that the apparent horizon of a black hole at the time
of formation cannot be bigger than the size of the cosmological horizon.
This upper bound was indeed shown to correspond to the mass of large
supercritical black holes in dust. For the case of a radiation, this
relation was only suggested by the behavior of $M_{bh}(R)$ for small
walls, extrapolated up to the critical size. With this assumption,
it was found that the peak of the distribution $f$ corresponds to
black holes of mass $M_{crit}\sim t_{\sigma}/G$, coming from the
collapse of a walls with $t_{H}(r_{i})\sim t_{\sigma}$. In this paper,
we will provide a more precise determination of $M_{bh}(R)$, which
should allow for a better comparison of the distribution $f$ with
observations.

Since the nucleation rate $\lambda$ is highly suppressed for $\sigma\gg H_{i}^{3}$,
and the semiclassical approximation breaks down for $\sigma\ll H_{i}^{3}$,
this leaves $\sigma\sim H_{i}^{3}$ as the preferred range of parameters.
For an inflationary Hubble rate in the interval $10^{14}\ {\rm GeV}\gtrsim H_{i}\gtrsim100\ {\rm GeV}$
the peak in $f$ lies within the range $10^{7}\ {\rm kg}\lesssim M_{crit}\lesssim10^{15}M_{\odot}$.
This leaves ample room for potentially observable effects on astrophysical
and cosmological scales. Black holes at the peak of the distribution
would form well within the radiation-dominated era, and therefore
it is important to explore this case in greater detail. This will
be the subject of the present paper.

In Section II we describe our model and the equations of motion. The
masses of subcritical and supercritical black holes as a function
of the initial radius are discussed in Sections III and IV, respectively.
Section V is devoted to a description of the evolution of the interior
of a supercritical wall filled with radiation. Section VI summarizes
our conclusions.

\section{Dynamics of spherical domain walls in an ambient FRW universe}

In this section we consider the equations of motion and boundary conditions
for a spherical domain wall embedded in an otherwise homogeneous FRW
universe, initially dominated by the energy density of a perfect fluid.

\subsection{Model and equations of motion}

We will be interested in a spherically symmetric situation, where
the metric can be written in the form 
\begin{equation}
ds^{2}=-A(r,t)dt^{2}+B(r,t)^{2}dr^{2}+R(r,t)^{2}d\Omega^{2}.\label{metric}
\end{equation}
Here $R(r,t)$ is the area radius, which we assume to vanish at the
origin of spherical coordinates, $R(0,t)=0$, and $d\Omega^{2}\equiv d\theta^{2}+\sin^{2}\theta d\varphi^{2}$.
There is some freedom in choosing the lapse function $A(r,t)$, and
in what follows we set $A(r,t)=1$, which completely fixes the gauge.
This is called geodesic slicing because observers at fixed spatial
coordinates experience free-fall \cite{numerical R}. 

A domain wall can be described by a scalar field $\phi(r,t)$ with
a double-well self-interaction potential $V(\phi)$. The Lagrangian
density is given by 
\begin{equation}
\mathcal{L}_{\phi}=\frac{1}{2}(\partial_{\alpha}\phi)(\partial^{\alpha}\phi)-V(\phi),
\end{equation}
and the corresponding energy-momentum tensor takes the form 
\begin{equation}
T_{\nu}^{(\phi)\mu}=\partial^{\mu}\phi\partial_{\nu}\phi-\delta_{\nu}^{\mu}\mathcal{L}_{\phi}.
\end{equation}
The background in which the domain wall is embedded is initially dominated
by a perfect fluid with energy-momentum tensor $T_{\nu}^{(f)\mu}=\left(\rho+p\right)u^{\mu}u_{\nu}+\delta_{\nu}^{\mu}p$,
where $\rho(r,t)$ and $p(r,t)=w\rho(r,t)$ are respectively the energy
density and pressure measured in the fluid frame ($w=0$ for dust
and $w=\frac{1}{3}$ for radiation). The fluid's 4-velocity can be
written in the form 
\begin{equation}
u^{\mu}(r,t)=\left(\frac{1}{\sqrt{1-v^{2}}},\frac{v}{B\sqrt{1-v^{2}}},0,0\right)
\end{equation}
where $v(r,t)$ is the fluid's 3-velocity relative to the comoving
coordinate $r$. We shall assume that there is only gravitational
interaction between the scalar field and the fluid, so that the total
energy-momentum tensor is simply given by $T_{\nu}^{(\phi)\mu}+T_{\nu}^{(f)\mu}$.

Our goal is to solve the equations of motion in order to find $B(r,t),R(r,t),\phi(r,t),\rho(r,t)$
and $v(r,t)$. Let $H_{i}$ be the Hubble parameter at the end of
inflation. By the following replacements 
\begin{equation}
r\to\frac{r}{H_{i}},\ t\to\frac{t}{H_{i}},\ \rho\to M_{Pl}^{2}H_{i}^{2}\rho,\ \phi\to M_{Pl}\phi,\ V\to M_{Pl}^{2}H_{i}^{2}V,\label{units}
\end{equation}
where $M_{Pl}=1/\sqrt{G},$ all variables become dimensionless. For
instance, $t_{H}$ becomes $2r_{i}^{3}/3$ for dust and $r_{i}^{2}/2$
for radiation. In what follows we use the dimensionless variables.

Following \cite{Cho,Bloomfield} we introduce 
\begin{equation}
U\equiv\dot{R},\ \Gamma\equiv\frac{R^{\prime}}{B},\ K\equiv\frac{\dot{B}}{B}+\frac{2\dot{R}}{R},
\end{equation}
where $\dot{}\equiv{\partial}/{\partial t}$ and $^{\prime}\equiv{\partial}/{\partial r}$.
The evolution equations then take the form 
\begin{equation}
\dot{K}=-\left(K-\frac{2U}{R}\right)^{2}-2\left(\frac{U}{R}\right)^{2}-4\pi(T_{00}+T_{\ 1}^{1}+2T_{\ 2}^{2})\label{K}
\end{equation}
\begin{equation}
\dot{U}=-\frac{1-\Gamma^{2}+U^{2}}{2R}-4\pi RT_{\ 1}^{1}\label{U}
\end{equation}
\begin{equation}
\dot{\Gamma}=-\frac{4\pi RT_{\ 1}^{0}}{B}\label{G}
\end{equation}
\begin{equation}
\dot{\rho}=\frac{(1+w)\rho}{1-wv^{2}}\left[v^{2}\left(K-\frac{2U}{R}\right)-K-\frac{2v\Gamma}{R}-\frac{v^{\prime}}{B}\right]-\frac{1-w}{1-wv^{2}}\frac{\rho^{\prime}v}{B}\label{rho}
\end{equation}
\begin{equation}
\dot{v}=\frac{(1-v^{2})v}{1-wv^{2}}\left[-\left(K-\frac{2U}{R}\right)+wK+\frac{2wv\Gamma}{R}-\frac{(1-v^{2})w}{(1+w)v}\frac{\rho^{\prime}}{\rho B}\right]-\frac{1-w}{1-wv^{2}}\frac{v^{\prime}v}{B}\label{v}
\end{equation}
\begin{equation}
\ddot{\phi}=-K\dot{\phi}+\frac{1}{BR^{2}}\left(\frac{R^{2}}{B}\phi^{\prime}\right)^{\prime}-\partial_{\phi}V\label{phi}
\end{equation}
\begin{equation}
\dot{B}=B\left(K-\frac{2U}{R}\right)\label{dK}
\end{equation}
\begin{equation}
\dot{R}=U\label{dU}
\end{equation}
where 
\begin{equation}
T_{00}=\frac{1+wv^{2}}{1-v^{2}}\rho+\frac{1}{2}\dot{\phi}^{2}+\frac{1}{2}\phi^{\prime2}+V(\phi)
\end{equation}
\begin{equation}
T_{\ 1}^{1}=\frac{w+v^{2}}{1-v^{2}}\rho+\frac{1}{2}\dot{\phi}^{2}+\frac{1}{2}\phi^{\prime2}-V(\phi)
\end{equation}
\begin{equation}
T_{\ 2}^{2}=w\rho+\frac{1}{2}\dot{\phi}^{2}-\frac{1}{2}\phi^{\prime2}-V(\phi)
\end{equation}
\begin{equation}
T_{\ 1}^{0}=\dot{\phi}\phi^{\prime}-\frac{1+w}{1-v^{2}}\rho vB
\end{equation}
Eq. (\ref{K}), (\ref{U}) and (\ref{G}) follow from Einstein equations,
while Eqs. (\ref{rho}) and (\ref{v}) follow from the conservation
of the fluid energy-momentum tensor, and Eq. (\ref{phi}) is the scalar
field equation. As for Eqs. (\ref{dK}) and (\ref{dU}), they are
satisfied by definition of $K$ and $U$ respectively. As we discuss
below, we use a combination of the Hamiltonian and the momentum constraint
in order to check for numerical accuracy {[}see Eq. (\ref{check}){]}.

A common technical difficulty in simulating a domain wall over a large
dynamical range is that the spacetime around the wall is expanding.
Hence, in comoving coordinates, the wall thickness is decreasing,
and more precision is needed to describe its sharper profile. To deal
with this problem, we adopt a conventional approach called adaptive
mesh refinement (AMR) \cite{AMR}. A detailed recent discussion of
this scheme can be found, for example, in Ref. \cite{BubbleCollision}.
The strategy is to assign a sufficient number of grid points to the
wall as the comoving thickness decreases, while the grid density far
away from the wall remains unaltered for the sake of simulation efficiency.
We require at least $\sim20$ grid points across the wall. To calculate
the derivatives on such a non-uniform grid, we use the method given
in Ref. \cite{derivative}.

The difficulty becomes particularly acute when the wall starts inflating
due to its self-gravity, since then $R$ grows exponentially in proper
time, and the comoving thickness of the wall shrinks exponentially.
As a result, mesh refinement becomes inefficient for supercritical
walls after the time $t_{\sigma}$. We will deal with this issue in
Section IV.

\subsection{Initial Conditions}

For convenience, we choose the so-called sine-Gordon self-interaction
potential for the scalar field, 
\begin{equation}
V(\phi)=\lambda\cos^{2}\left(\frac{\phi}{2\eta}-\frac{\pi}{2}\right).
\end{equation}
The corresponding field equation has an analytic planar wall solution
in flat spacetime, given by 
\begin{equation}
\phi_{w}(x)=-4\eta\arctan(e^{\frac{x}{\eta}\sqrt{\frac{\lambda}{2}}})+2\eta\pi.
\end{equation}
We use this as the initial profile of the scalar field, in the form
\begin{equation}
\phi(r,t_{i})=\phi_{w}(r-r_{i}).
\end{equation}
In dimensionless units (\ref{units}), the initial energy density
at the time $t_{i}$ in a FRW universe is given by $\rho_{i}=(3/8\pi)$.
The energy density in the core of the domain wall is assumed to be
subdominant at that time, and so we require $\lambda\ll1$. The thickness
of the wall is given by $\delta\sim{\eta}/{\sqrt{\lambda}}$ and should
be smaller than the initial Hubble radius, $\delta\lesssim1$. It
follows that we must choose $\eta\lesssim\sqrt{\lambda}\lesssim1$.
With this, the dimensionless surface tension $\sigma\sim\eta\sqrt{\lambda}\lesssim1$
will also be small, and $t_{\sigma}=(2\pi\sigma)^{-1}\gtrsim1$. For
supercritical walls we need $t_{\sigma}\lesssim t_{H}$. In particular,
in the marginal case where $\sigma\sim1$, all walls with initial
size much larger than the Hubble radius, $r_{i}\gg1$, will be supercritical.

For the metric components and fluid velocity we assume the initial
conditions $B(r,t_{i})=1,$ $R(r,t_{i})=r$ and $v(r,t_{i})=0.$ Further
we assume that $\dot{\phi}(r,t_{i})=0$, and that the small initial
energy density of the wall is compensated for by a deficit in the
energy density of the background fluid, so $\rho(r,t_{i})=(3/8\pi)-(1/2)\phi^{\prime2}(r,t_{i}\text{)}-V(\phi)$.
The initial values of $K$ and $U$ can be determined from $G_{00}=8\pi GT_{00}$
and $G_{01}=8\pi GT_{01}$, and we find $K(r,t_{i})=3$ and $U(r,t_{i})=r$,
as in an FRW universe.

\subsection{Boundary Conditions}

We also need to set boundary conditions at $r=0.$ Obviously $R(0,t)=0$
and $U(0,t)=0.$ By regularity, $\rho^{\prime}(0,t)=0,$ $v^{\prime}(0,t)=0$,
$B^{\prime}(0,t)=0$ and $B(0,t)=\lim_{r\rightarrow0}R(r,t)/r$, which
gives $K(0,t)=\lim_{r\rightarrow0}3U(r,t)\text{/}R(r,t)$ and $\Gamma(0,t)=1$.

Note that there are no physical restrictions at the outer boundary
in our computational domain, because derivatives there do not necessarily
vanish (as they do in an FRW region). However, the only physical perturbation
comes from the rarefaction waves produced near the wall, and as long
as we have a sufficiently large computational domain, we don't need
to worry about reflected waves. Therefore, to evolve the outer boundary,
we don't specify any boundary condition, but simply use one-sided
derivatives \cite{numerical R}. This works well enough for our purposes.

\subsection{Expansions and Misner-Sharp Mass}

Consider a spherical $r={\rm const.}$ surface and a congruence $k^{\mu}$
of null geodesics orthogonal to it, with expansion given by $\Theta\equiv\nabla_{\mu}k^{\mu}$.
We denote by $\Theta_{out}$ and $\Theta_{in}$ the expansion of outgoing
and ingoing radial null geodesics, respectively. A surface is trapped
if both expansions are negative, and anti-trapped if both are positive.
The expansion of null congruence is useful in characterizing black
holes locally (see e.g. \cite{numerical R}). In our coordinate system,
we have 
\begin{equation}
\Theta_{out}\propto\frac{U+\Gamma}{R},\ \Theta_{in}\propto\frac{U-\Gamma}{R}.\label{theta}
\end{equation}
In a flat FRW universe, 
\begin{equation}
\Theta_{out}\propto H+\frac{1}{R},\ \Theta_{in}\propto H-\frac{1}{R}\label{FRW-1}
\end{equation}
which should be satisfied in regions far from the wall.

We may characterize the quasi-local mass of an object by adopting
the concept of Misner-Sharp energy \cite{MS,hayward}, which for a
general metric $ds^{2}=h_{ab}dx^{a}dx^{b}+R^{2}d\Omega^{2}$ is defined
as 
\begin{equation}
M=\frac{R}{2}(1-h^{ab}\partial_{a}R\partial_{b}R).
\end{equation}
and in our coordinates is given by 
\begin{equation}
M=\frac{R}{2}(1-\Gamma^{2}+U^{2}).
\end{equation}
For a perfect fluid, the energy-momentum tensor is diagonal. In this
case, 
\begin{equation}
8\pi T_{00}=G_{00}=\frac{2M^{\prime}}{R^{2}R^{\prime}},\label{MS}
\end{equation}
which gives $M=4\pi\int T_{00}R^{2}dR$. So it is natural to interpret
$M$ as ``mass''. For an FRW universe, $M$ is the total mass of
matter enclosed by a sphere with radius $R$. In the case of Schwarzschild
spacetime, $M$ is the mass constant in the metric. However, in our
case, $T_{\ \nu}^{\mu}$ is not diagonal due to the scalar field.
Instead of Eq. (\ref{MS}), now we have 
\begin{equation}
M^{\prime}=4\pi R^{2}\left(R^{\prime}T_{00}-\dot{R}T_{01}\right).\label{check}
\end{equation}
We will use this equation as a check for numerical accuracy.

An important point for us about the Misner-Sharp mass is that it is
not necessarily monotonic in $r$, because $R$ is not necessarily
monotonic in $r$. In supercritical cases, due to the existence of
a wormhole, there is a local minimum of $R$ and the mass $M$ could
decrease with increasing $r$.

The apparent horizon of a spherically symmetric black hole is a marginally
trapped surface, where $\Theta_{out}=0$ and $\Theta_{in}<0.$ By
definition we have $R=2M$ at the apparent horizon. We also define
the white hole horizon and cosmological horizon to be a marginally
anti-trapped hypersurface where $\Theta_{in}=0$ and $\Theta_{out}>0.$
A spherical surface is anti-trapped (with $\Theta_{in}>0$ and $\Theta_{out}>0$)
within the white hole horizon, and is normal (with $\Theta_{in}<0$
and $\Theta_{out}>0$) if it lies between the white hole horizon and
the cosmological horizon. Note that by Eq. (\ref{FRW-1}), the cosmological
horizon and the Hubble radius coincide in a flat FRW universe.

\section{subcritical Wall ($t_{\sigma}\gtrsim t_{H}$)}

\subsection{Equation of Motion}

When $t_{\sigma}\gg t_{H}$, the wall's gravitational effect can be
completely neglected until the wall is about to collapse. Hence, the
dynamics is well approximated by that of a test wall moving in the
FRW background. In this case the wall's equation of motion is given
by \cite{Garriga} 
\begin{equation}
\ddot{r}+(4-3a^{2}\dot{r}^{2})H\dot{r}+\frac{2}{a^{2}r}(1-a^{2}\dot{r}^{2})=0\label{3}
\end{equation}
where $r$ is the wall's comoving radius, $a$ is the scale factor,
and $H\equiv{\dot{a}}/{a}$ is the Hubble parameter. When we get closer
to the critical regime, the wall's gravitational effect should be
taken into account, and a similar equation can be obtained (see Appendix),
\begin{equation}
\ddot{r}+(4-3a^{2}\dot{r}^{2})H\dot{r}+\frac{2}{a^{2}r}(1-a^{2}\dot{r}^{2})=6\pi\sigma\frac{(1-a^{2}\dot{r}^{2})^{\frac{3}{2}}}{a}\label{4}
\end{equation}
This equation is derived by assuming that the FRW spacetime inside
the wall is unperturbed. We see that the only difference between Eq.
(\ref{3}) and Eq. (\ref{4}) is the right hand side, which is a term
related to the surface tension. It is also shown in Appendix that
the time $t_{BH}$ when the apparent horizon first appears is given
by 
\begin{equation}
\left(H+\frac{1}{ar}\right)\sqrt{\frac{1+a\dot{r}}{1-a\dot{r}}}=4\pi\sigma\label{5}
\end{equation}
It turns out Eqs. (\ref{4}) and (\ref{5}) agree very well with our
numerical solution of the full equations of motion.

\begin{figure}
\includegraphics[scale=0.14]{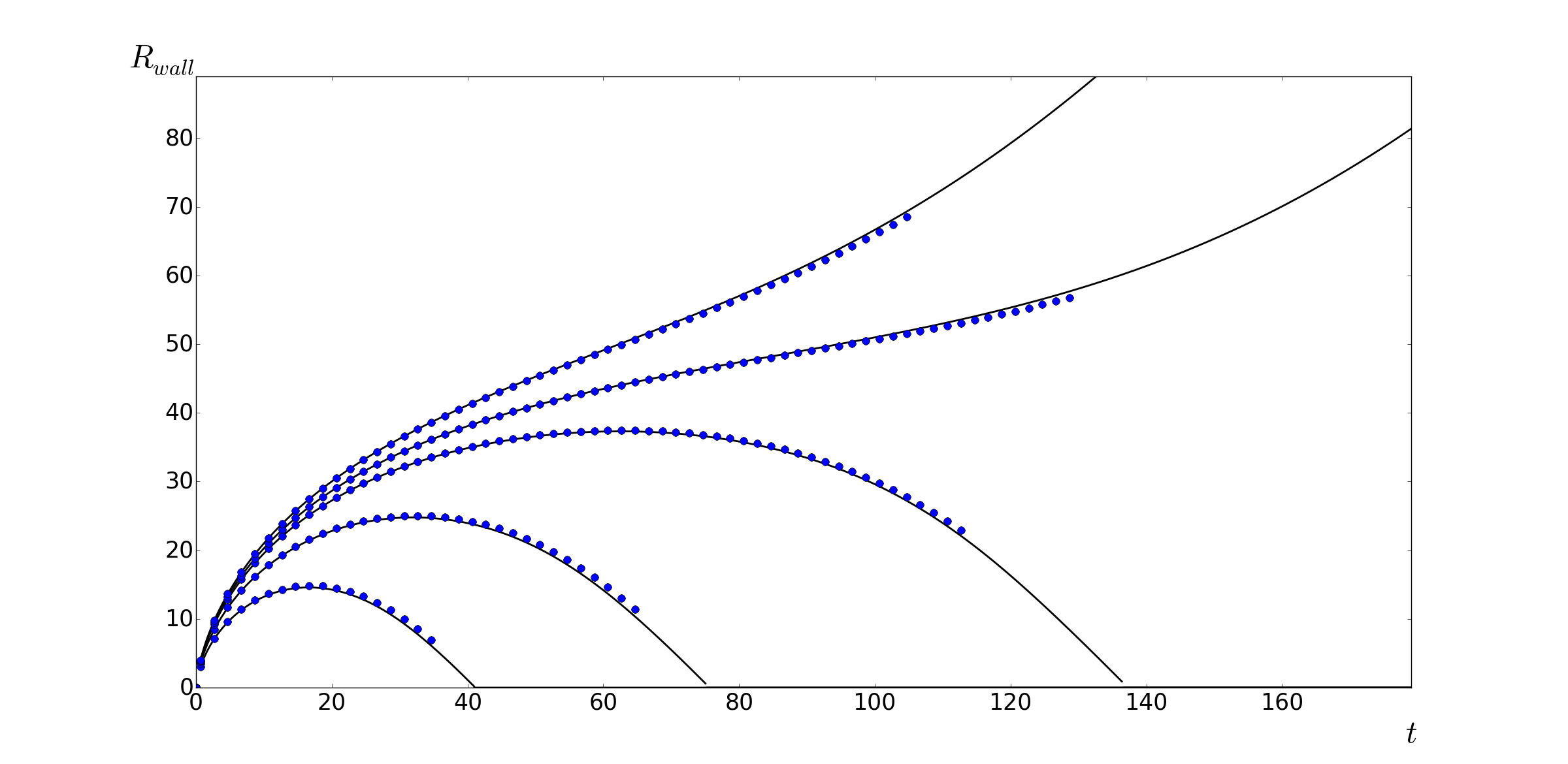}\includegraphics[scale=0.14]{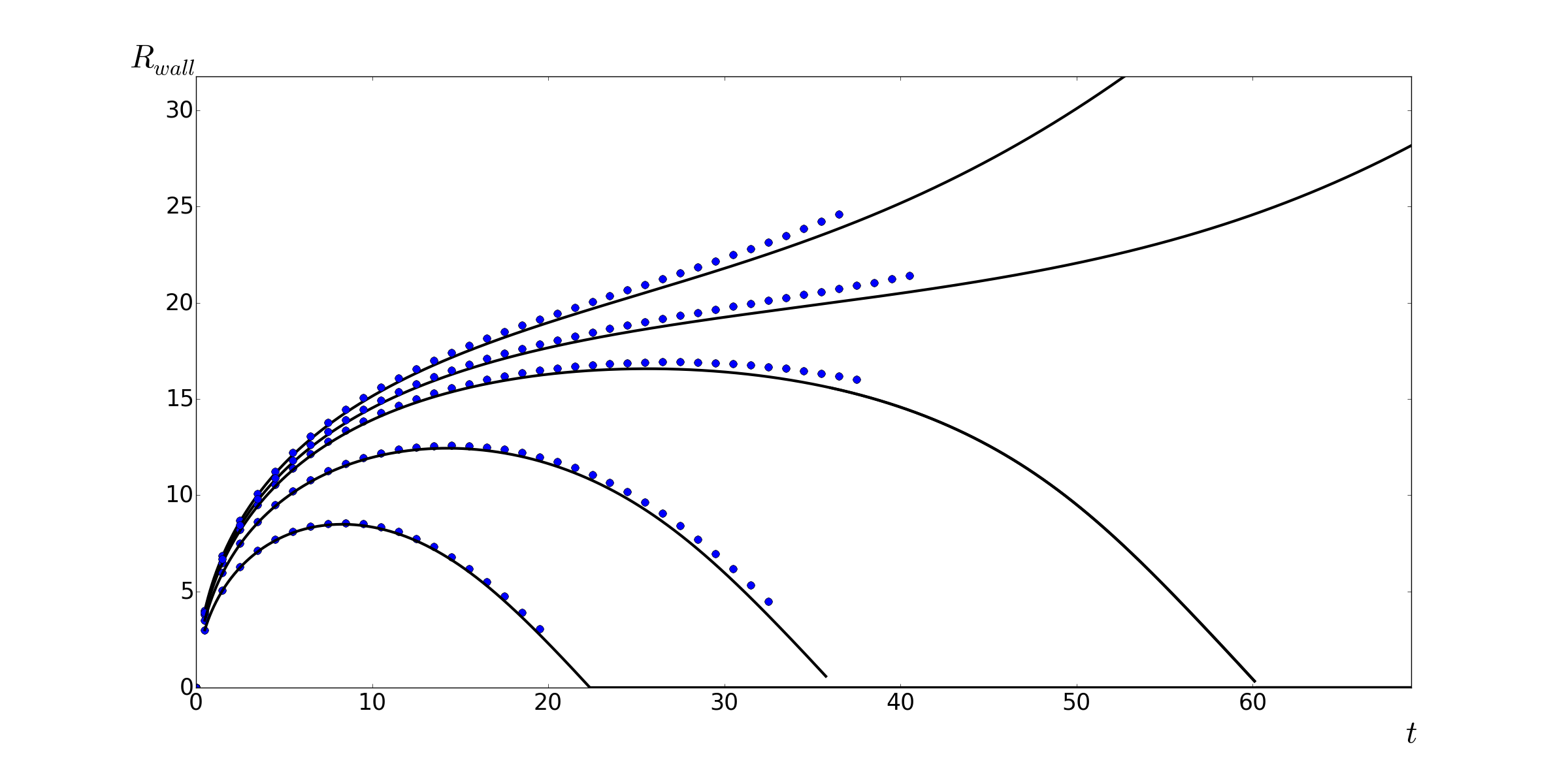}\protect\caption{\label{R-t}Trajectories of five walls in dust background (left panel,
with $t_{\sigma}\approx74$) and radiation background (right panel,
with $t_{\sigma}\approx30$), with $r_{i}=3,3.5,3.8,3.9,4$ from the
bottom up. $R_{wall}=ar$ is the wall's area radius. Solid curves
are obtained from Eq. (\ref{4}), and dots are from simulations. We
can see that Eq. (\ref{4}) works very well in the case of dust background
even for walls near the critical regime, where they start inflating.}
\end{figure}

For dust background they work even near the critical regime where
$t_{\sigma}\sim t_{H}$ (see Fig. \ref{R-t}). In the case of radiation,
the propagation of rarefaction waves produced near the wall influences
the interior FRW region, but Eq. (\ref{4}) is satisfied to a large
extent. Eqs.~(\ref{4}) and (\ref{5}) can easily be solved numerically,
and instead of running full simulations, it is much more convenient
to use them to obtain the initial black hole mass. This is simply
given by $M_{BHi}=(1/2)a(t_{BH})r(t_{BH})$, where $r(t_{BH})$ is
the location where the apparent horizon first forms.

\subsection{Initial BH}

It would be useful to have an estimate of the black hole mass at the
time of formation, $M_{BHi}$, from the initial conditions. Ref. \cite{Ipser}
gives an expression for the black hole mass formed by domain wall
collapse in vacuum, $M_{IS}=4\pi\sigma R_{m}^{2}(1-2\pi\sigma R_{m}),$
where $R_{m}$ is the maximum area radius of the wall and can be determined
by Eq. (\ref{4}). Let $M_{BHi}=C_{IS}M_{IS}$ with $C_{IS}$ a coefficient.
We find that solving Eq. (\ref{4}) and (\ref{5}) gives $C_{IS}\approx1.1$
for both dust and radiation.

Now let $R_{H}={1/H(t_{H})}$, which is simply given by $r_{i}^{3}$
for dust and $r_{i}^{2}$ for radiation. Physically, we expect $R_{m}$
to be of the same order as $R_{H}$, since the wall comes to a halt
and then collapses due to its tension once it falls within the cosmological
horizon. When $t_{\sigma}\gg t_{H},$ we may assume $M_{BHi}=4\pi\sigma CR_{H}^{2}$,
where $C$ is a coefficient. By solving Eq. (\ref{3}), it was found
in Ref. \cite{Garriga} that $C_{dust}\approx0.15$ and $C_{radiation}\approx0.62$.
Solving Eq. (\ref{4}) certainly gives similar results.

\subsection{Final BH}

A singularity arises soon after the apparent horizon forms. A well
known drawback of the geodesic slicing that we are using is that it
inevitably hits the singularity \cite{numerical R}. To avoid the
breakdown of the simulation, we cut off the black hole region by ignoring
what happens inside the apparent horizon, and use one-sided boundary
conditions there. Since the apparent horizon lies inside the event
horizon, it is safe to do so. Such black hole excision method \cite{numerical R}
allows us to evolve the spacetime to a much later time and observe
the mass accretion.

\begin{figure}
\includegraphics[scale=0.2]{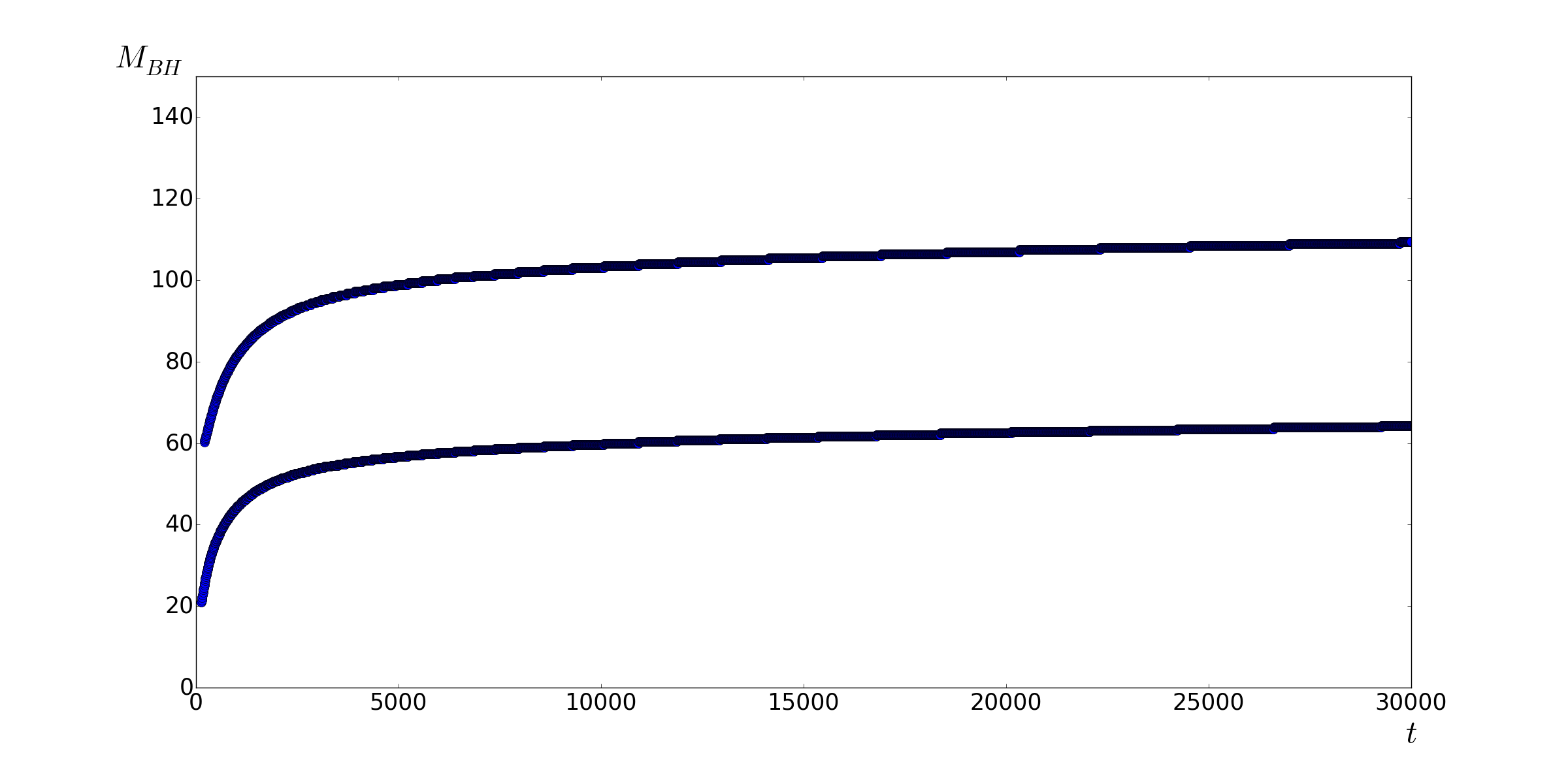}\protect\caption{\label{M-t-1-2}$M_{BH}$ as a function of time for two subcritical
walls in dust background with $r_{i}=5$ and $6$, and $t_{\sigma}\approx300.$
After the black hole is formed, its mass increases and converges to
$M_{BHf}\approx(1/2)r_{i}^{3}$. }
\end{figure}

\begin{figure}
\includegraphics[scale=0.13]{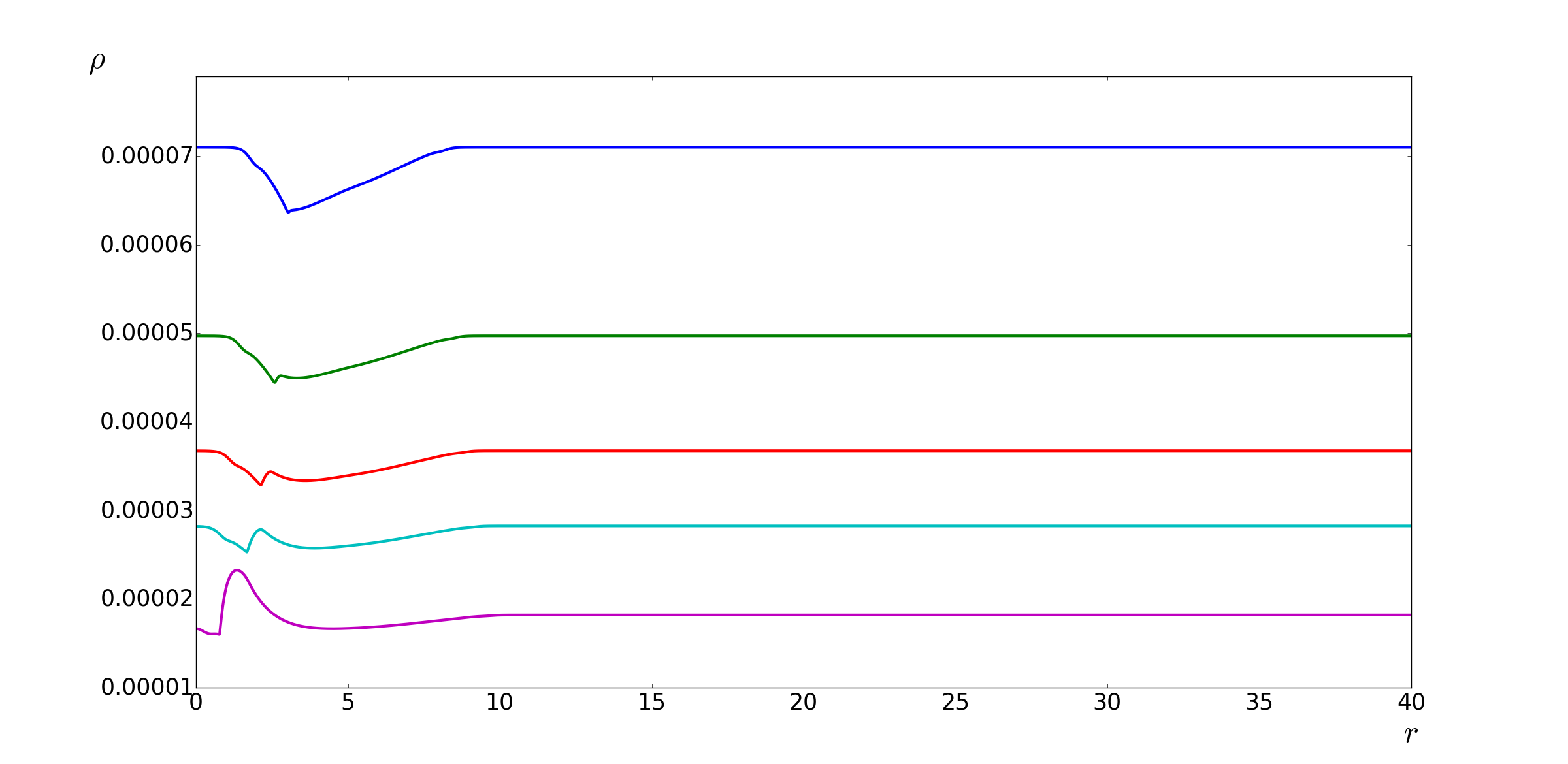}\includegraphics[scale=0.14]{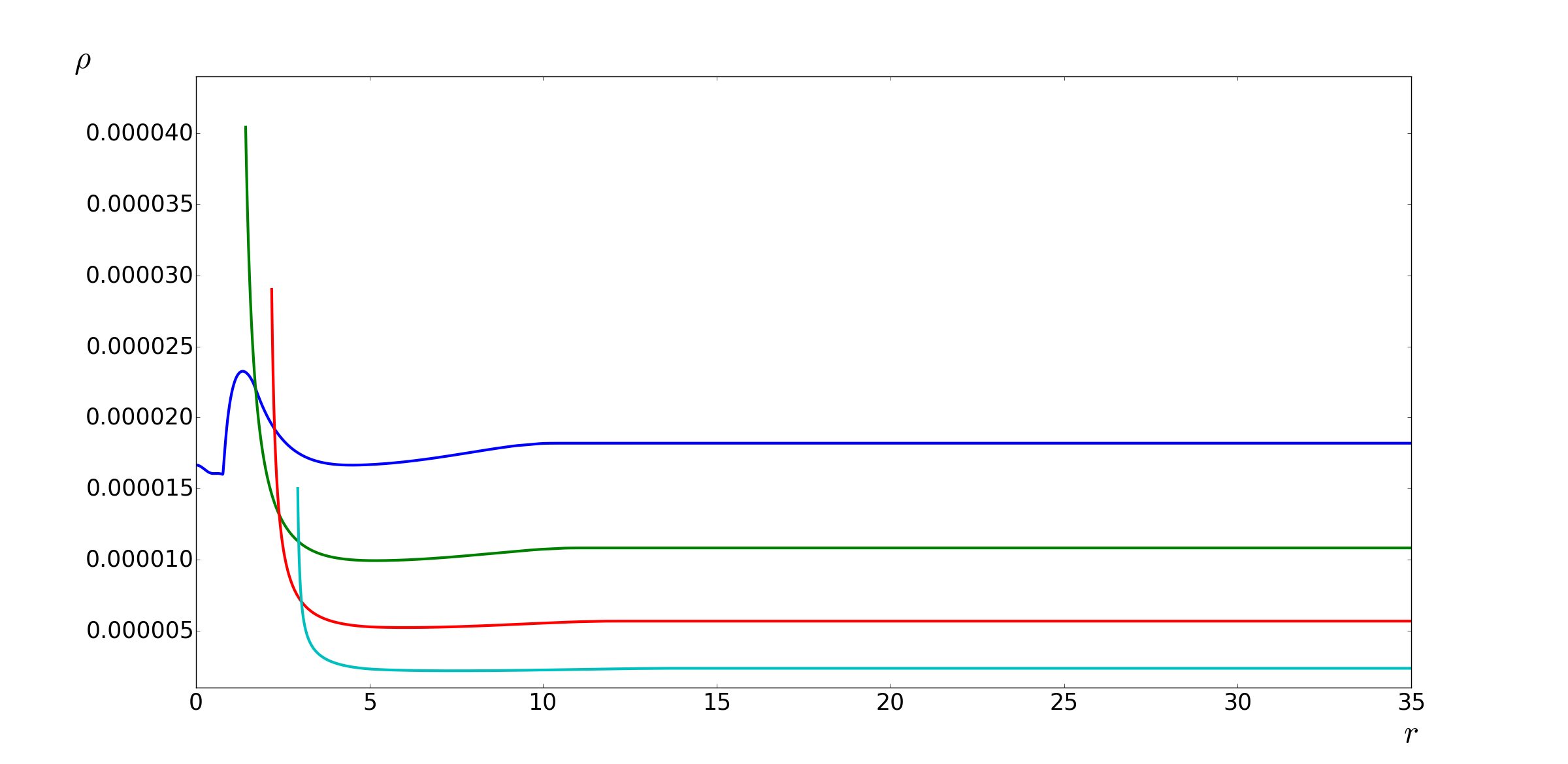}\protect\caption{\label{rho-r}Evolution of the radiation energy density distribution
in the case of a subcritical wall with $r_{i}=5$ and $t_{\sigma}\approx300$.
The left panel shows $\rho(r)$ at time $t=20.5,24.5,28.5,32.5,40.5$
from the top. The wall (which is not shown in the plot) is at the
position where $\rho$ has the minimum value. We can see that rarefaction
waves (which will be discussed later) are produced and propagate away
from the wall. Meanwhile in the exterior region the energy density
decreases as in an FRW universe. A ``bump'' develops outside the
wall and will be cut off later because it will be within the black
hole apparent horizon. The right panel shows $\rho(r)$ at time $t=40.5,52.5,72.5,112.5$
from the top. The black hole (which is excised for the last three
moments in the plot) is not surrounded by an empty layer as it is
in a dust universe. We can see that the energy deficit between the
apparent horizon and the unperturbed FRW region is smoothed out with
time.}
\end{figure}

The evolution of the black hole mass, $M_{BH}(t)$, is different for
dust and radiation. It was shown in \cite{Tanahashi} that, for a
dust background, all of the fluid between the apparent horizon and
the exterior FRW boundary (which is also the approximate comoving
location of the initial wall), gradually falls into the black hole,
while the exterior region is unaffected. Eventually, we have a black
hole enclosed by an empty region surrounded by a dust universe, where
the final black hole mass $M_{BHf}$ is simply the enclosed mass within
the initial wall, and is given by $M_{H}=(1/2)r_{i}^{3}$ (see Fig.
\ref{M-t-1-2}).

\begin{figure}
\includegraphics[scale=0.25]{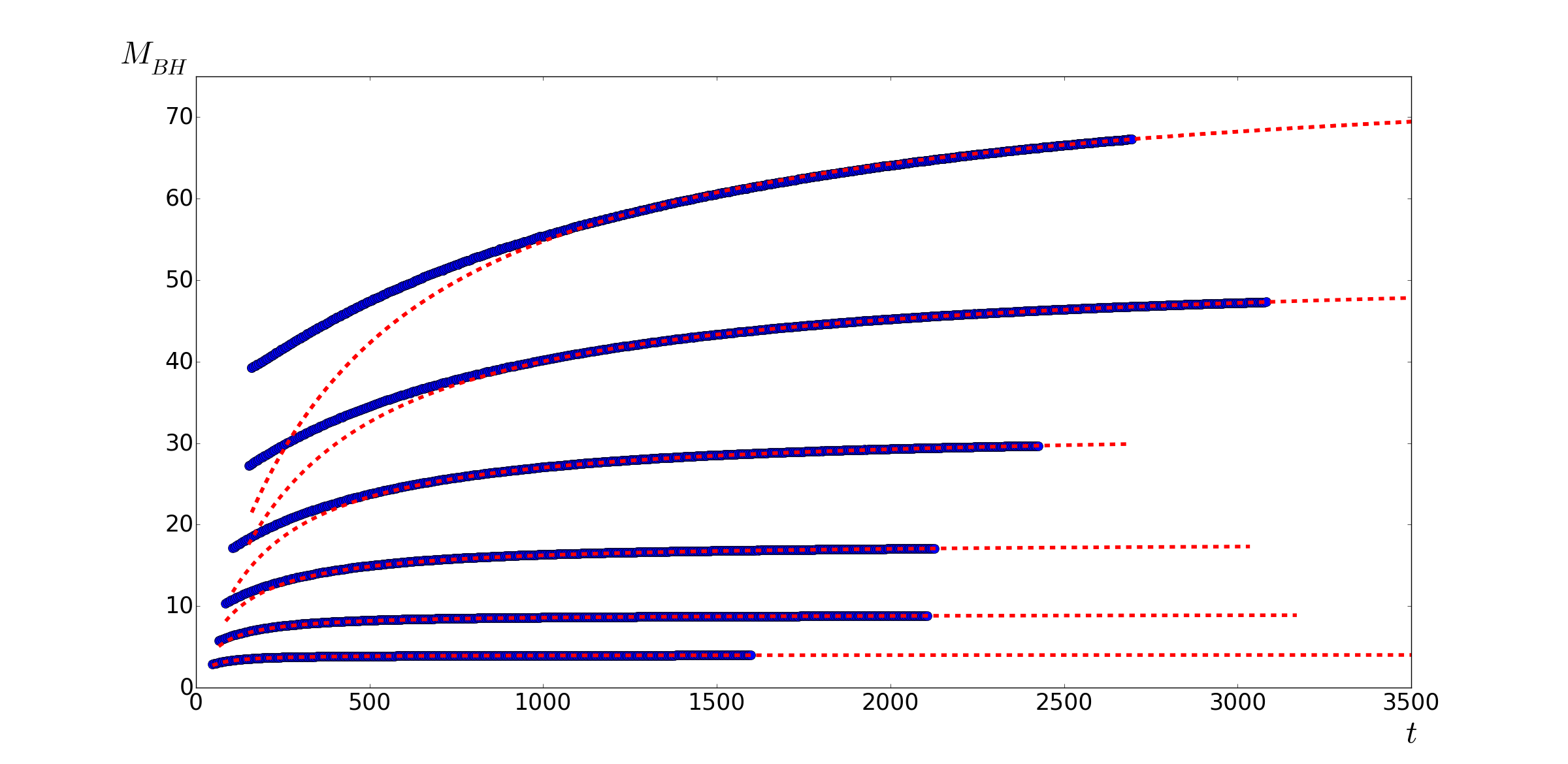}\protect\caption{\label{M-t-1}$M_{BH}$ as a function of time for six subcritical
walls in the background of radiation with the same surface tension
($t_{\sigma}\approx300$) but different radii. $r_{i}=5,6,7,8,9\ \mbox{and }10$
from the bottom. Blue curves are from simulations, and dashed red
curves are from Eq. (\ref{accretion}). For $r_{i}=5$, $\frac{M_{BHf}}{M_{BHi}}\approx1.5$;
for $r_{i}=6$, $\frac{M_{BHf}}{M_{BHi}}\approx1.6$; for $r_{i}=7,$
$\frac{M_{BHf}}{M_{BHi}}\approx1.8$; for $r_{i}=8,$ $\frac{M_{BHf}}{M_{BHi}}\approx1.9$;
for $r_{i}=9,$ $\frac{M_{BHf}}{M_{BHi}}\approx2.0$; for $r_{i}=10,$
$\frac{M_{BHf}}{M_{BHi}}\approx2.0$. The ratio increases to $\sim2$
as we approach the critical regime. }
\end{figure}

However, in a radiation-dominated universe, fluid keeps being pushed
in and there will not be an empty layer outside the black hole horizon
(see Fig. \ref{rho-r}). The mass $M_{BH}(t)$ keeps growing as the
FRW energy density decreases (see Fig. \ref{M-t-1}).

The accretion of primordial black holes in the early universe was
first studied in Ref. \cite{Zeldovich}. If we assume that the radiation
energy density right outside the black hole horizon decreases as in
an FRW universe, the accretion rate of the black hole can be written
as 
\begin{equation}
\frac{dM_{BH}(t)}{dt}=4\pi FR_{BH}^{2}\rho(t)\label{Zeldovich}
\end{equation}
where $R_{BH}=2M_{BH}$ is the Schwarzschild radius, and the accretion
efficiency $F$ is expected to be a constant of order one \cite{Guedens,Nayak}.
The solution is 
\begin{equation}
M_{BH}(t)=\frac{1}{\frac{1}{M_{0}}+\frac{3}{2}F\left(\frac{1}{t}-\frac{1}{t_{0}}\right)}\label{accretion}
\end{equation}
where $M_{0}$ is the initial black hole mass and $t_{0}$ the time
when the black hole is formed.

\begin{figure}
\includegraphics[scale=0.2]{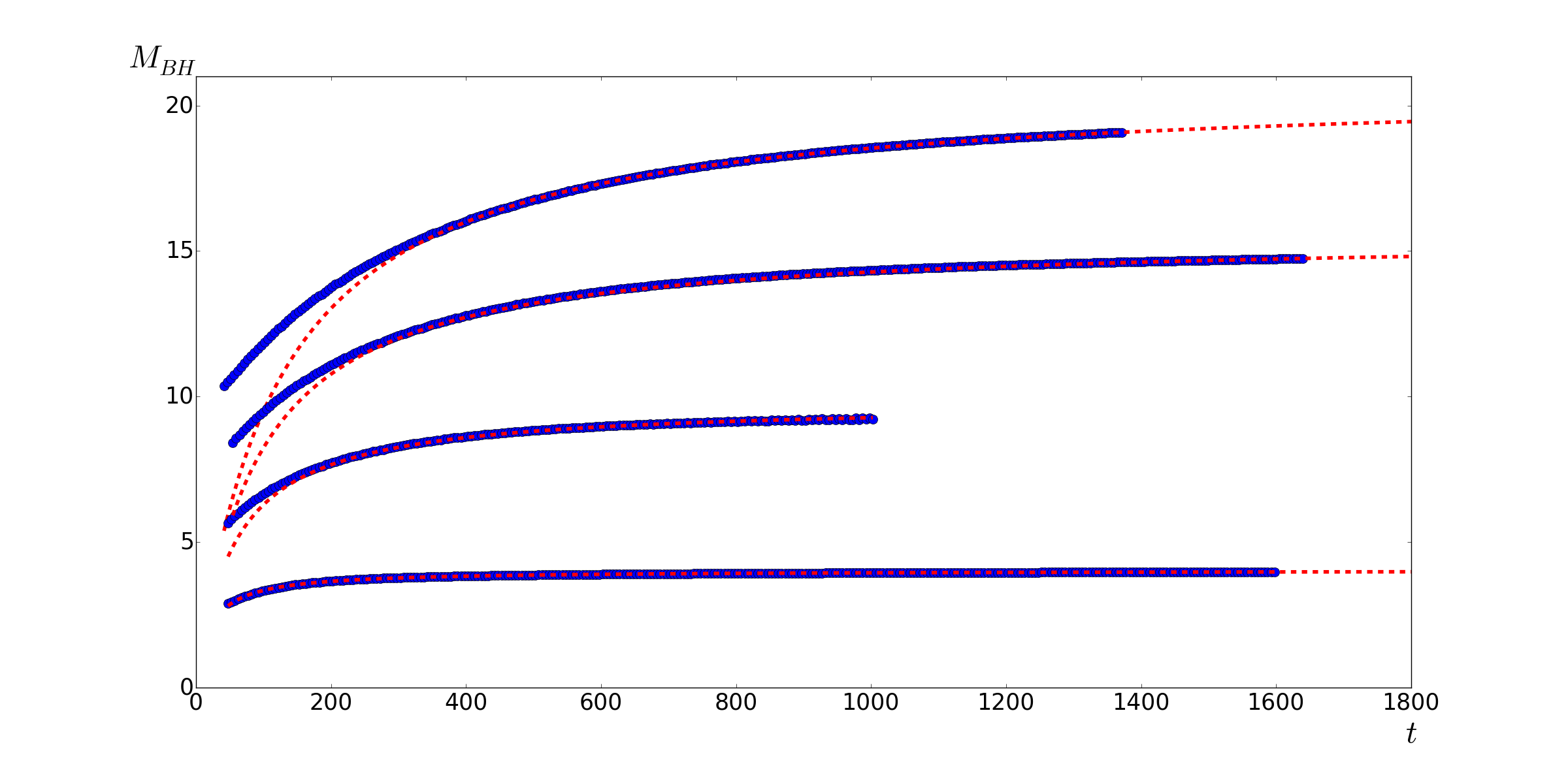}\protect\caption{\label{M-t-1-1}$M_{BH}$ as a function of time for four subcritical
walls in the background of radiation with the same $r_{i}(=5)$ but
different surface tension. $t_{\sigma}\approx200,100,50\ \mbox{and }25$
from the bottom. Blue curves are from simulations, and dashed red
curves are from Eq. (\ref{accretion}). For $t_{\sigma}=200$, $\frac{M_{BHf}}{M_{BHi}}\approx1.5$;
for $t_{\sigma}=100$, $\frac{M_{BHf}}{M_{BHi}}\approx1.8$; for $t_{\sigma}=50$,
$\frac{M_{BHf}}{M_{BHi}}\approx2.0$; for $t_{\sigma}=25$, $\frac{M_{BHf}}{M_{BHi}}\approx2.0$.
The ratio increases to $\sim2$ as we approach the critical regime.}
\end{figure}

In trying to fit our numerical results with this equation, we have
to take into account that the spacetime near the apparent horizon
is not FRW when the black hole is formed, so $M_{0}$ is not necessarily
$M_{BHi}$. Instead, we choose values for $M_{0}$ and $t_{0}$ at
a sufficiently late time. In the original and some subsequent papers
\cite{Bugaev,Majumdar}, $R_{BH}$ is replaced by an effective capture
radius of relativistic particles by the black hole, $R_{eff}=({3\sqrt{3}}/{4})R_{BH}$.
This gives $F\sim{27}/{16}$. Some argued \cite{Custodio} that $\rho$
should be replaced by $\rho+3p=2\rho$, to account for the difference
between the radiation energy density and the gravitational energy
density. Combining these two effects gives $F\approx3.4$. We find
that $F\approx3.8$ works very well for the data from all our simulations.
From Fig. \ref{M-t-1} we can see that, for relatively large walls,
the equation doesn't match the data right after the black hole formation,
but works perfectly at later times. With this in mind, we don't need
to run the simulation and wait until $M_{BH}(t)$ converges to a constant,
since as soon as Eq. (\ref{accretion}) is satisfied, we can immediately
infer the final mass: 
\begin{equation}
M_{BHf}=\frac{1}{\frac{1}{M_{0}}-\frac{5.7}{t_{0}}}.\label{accretion2}
\end{equation}
Here $M_{0}$ and $t_{0}$ can be the values at any sufficiently late
time. Our simulations suggest that the time it takes for the accretion to settle into the regime described by Eq. (\ref{accretion}) is about  $10-100\ M_{BHi}$.

Note that for relatively small walls, the black hole doesn't grow
much. By increasing $r_{i}$ or $\sigma$, the ratio $\frac{M_{BHf}}{M_{BHi}}$
increases until the critical regime is reached. In this regime, the
ratio saturates to $\frac{M_{BHf}}{M_{BHi}}\approx2$ (See captions
under Fig. \ref{M-t-1} and \ref{M-t-1-1}). In Section IV we will
see that the size of a supercritical black hole as measured by an
observer in the exterior asymptotically FRW region also doubles after
its formation.

\begin{figure}
\includegraphics[scale=0.2]{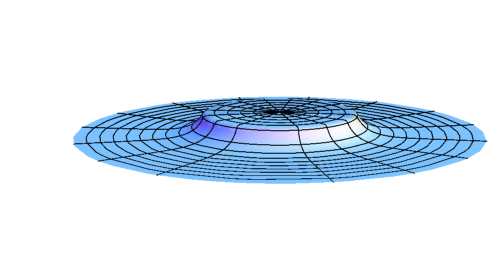}\includegraphics[scale=0.2]{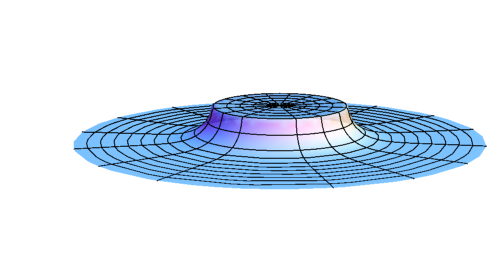}\includegraphics[scale=0.2]{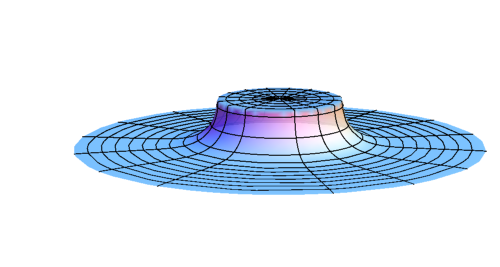}\includegraphics[scale=0.2]{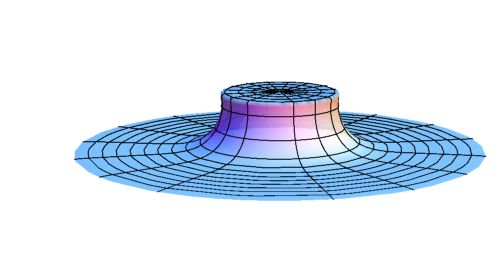}

\includegraphics[scale=0.2]{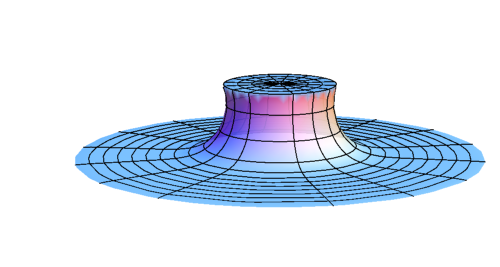}\includegraphics[scale=0.2]{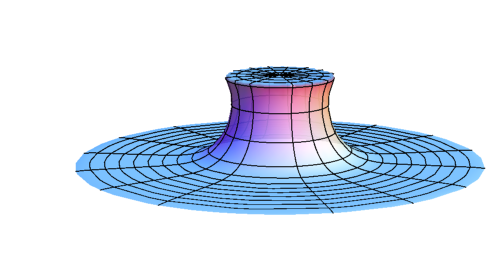}\includegraphics[scale=0.2]{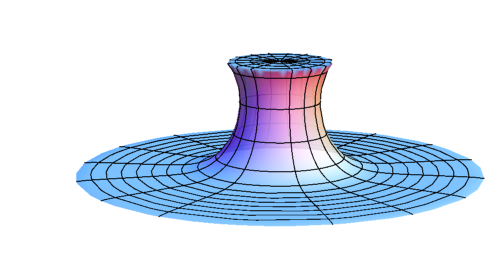}\includegraphics[scale=0.2]{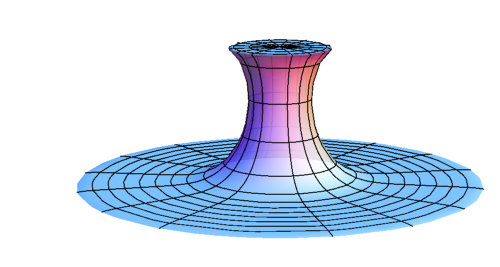}\protect\caption{\label{wormhole}Embedding diagrams showing how a wormhole develops
with time outside the wall. The flat-looking region at the top and
the bottom area represent the interior and exterior regions respectively.
The ring that encircles the top is where the wall is located. The
distance from the cap center along a longitude line is the proper
radius $d(r,t)\equiv\int_{0}^{r}B(\tilde{r},t)d\tilde{r}$, while
the radius of a latitude circle is the area radius $R$. The throat
will be pinched off when the black hole singularity is encountered.
These diagrams are not shown with the same scale. The radius of the
ring should grow exponentially. }
\end{figure}

\section{Supercritical Wall ($t_{\sigma}\lesssim t_{H}$)}

Gravitational collapse in the supercritical case is geometrically
rather interesting. Although the wall grows without bound, it does
so in a baby universe. First, a wormhole develops in the region between
the wall and the exterior FRW universe. The wormhole evolves into
a black hole interior region, bounded by two future black hole horizons,
one facing the asymptotic FRW universe, and the other facing the baby
universe (see Fig. \ref{wormhole}). The wormhole eventually ``pinches
off`` at the black hole singularity. This effectively disconnects
the baby universe from the exterior parent universe.

As shown in Ref. \cite{Garriga}, if we consider a thin wall in a
dust background, the wall repels the fluid nearby, and is subsequently
sandwiched by two Schwarzschild layers with the same mass parameter
$M_{BH}$. In the strongly supercritical limit, this mass is equal
to the mass of matter initially enclosed by the shell. In the outer
empty layer, there is a wormhole connecting two normal regions. From
either side of the wormhole throat a black hole can be observed. In
simulations, however, we find that the two masses ($M_{BHf}^{(in)}$
and $M_{BHf}^{(out)}$) are not exactly identical because more fluid
falls in from outside due to the initial wall thickness. More importantly,
in a radiation-dominated universe, there is also mass accretion, as
happened in the subcritical case. As we shall see, for a dust background
$M_{BHf}^{(in)}\approx M_{BHf}^{(out)}$, while for radiation $M_{BHf}^{(out)}\approx2M_{BHf}^{(in)}$.

\subsection{Spacetime Structure}

\begin{figure}
\includegraphics[scale=0.2]{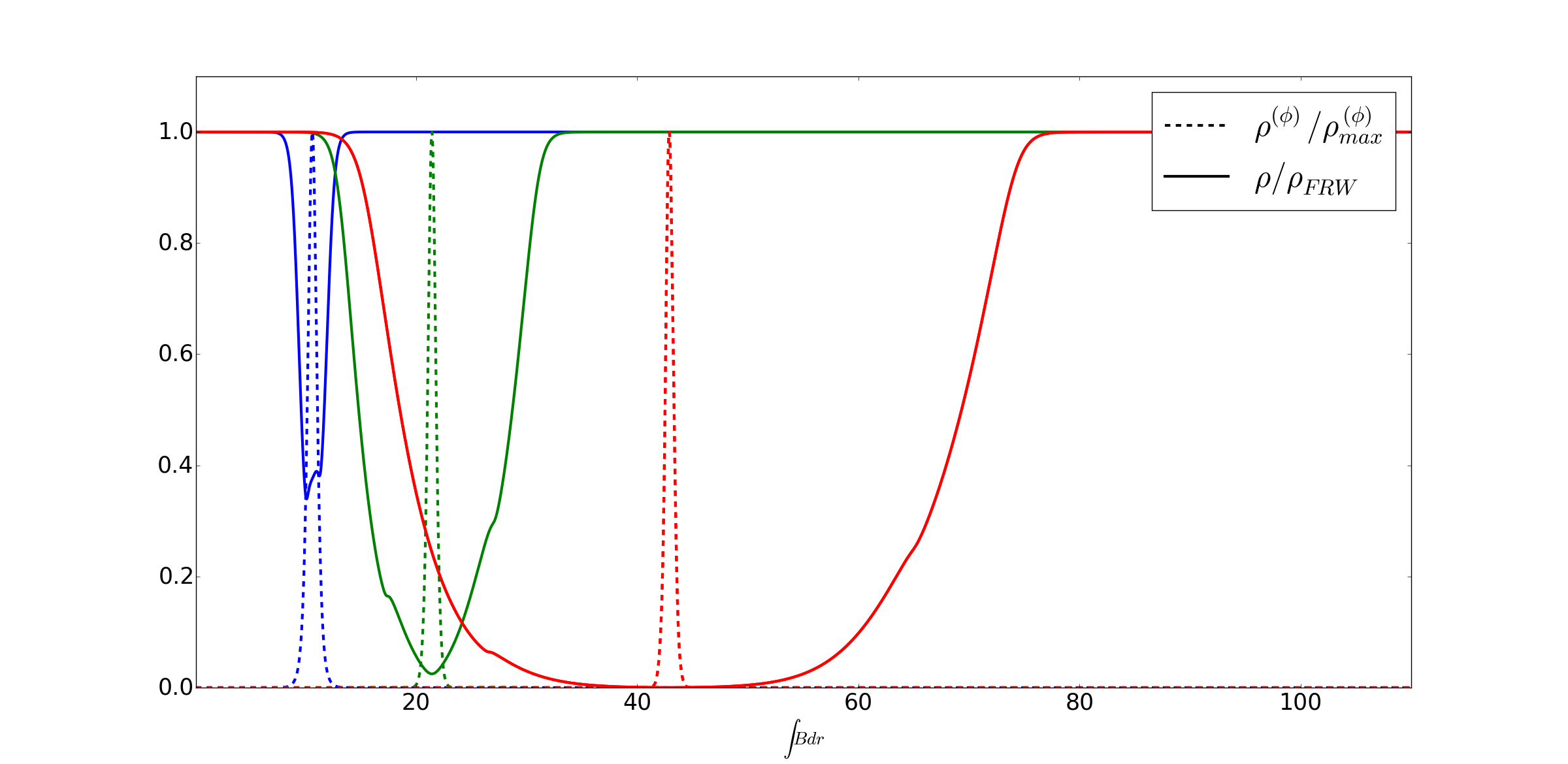}

\protect\caption{\label{proradius}This graph shows how the wall pushes radiation away.
Both fluid and field energy density have been rescaled so that the
largest value is 1. The blue curves are the earliest curves in this
set. The $x-$axis is the proper radius $d(r,t)\equiv\int_{0}^{r}B(\tilde{r},t)d\tilde{r}$
on a constant time hypersurface. The proper thickness of the wall
barely changes.}
\end{figure}

Using adaptive mesh refinement, we can evolve the domain wall for
some time. However, once the wall starts expanding exponentially,
more and more grid points need to be introduced to cover a wall that
keeps getting thinner in comoving coordinates. At that point the code
becomes computationally expensive. Fortunately, in this regime the
wall moves supersonically away from both the interior and the exterior
regions, and becomes detached from the fluid. Afterwards, the wall
remains in between two almost empty layers, where the energy density
of both the scalar field and the fluid is completely negligible compared
to the FRW density (Fig. \ref{proradius}). Since we are mostly interested
in the black hole as it is seen from the exterior FRW region, we removed
the wall and the region interior to the wall in some of the simulations.
The evolution of the interior region will be considered separately
in Section \ref{fireball}.

\begin{figure}
\includegraphics[scale=0.2]{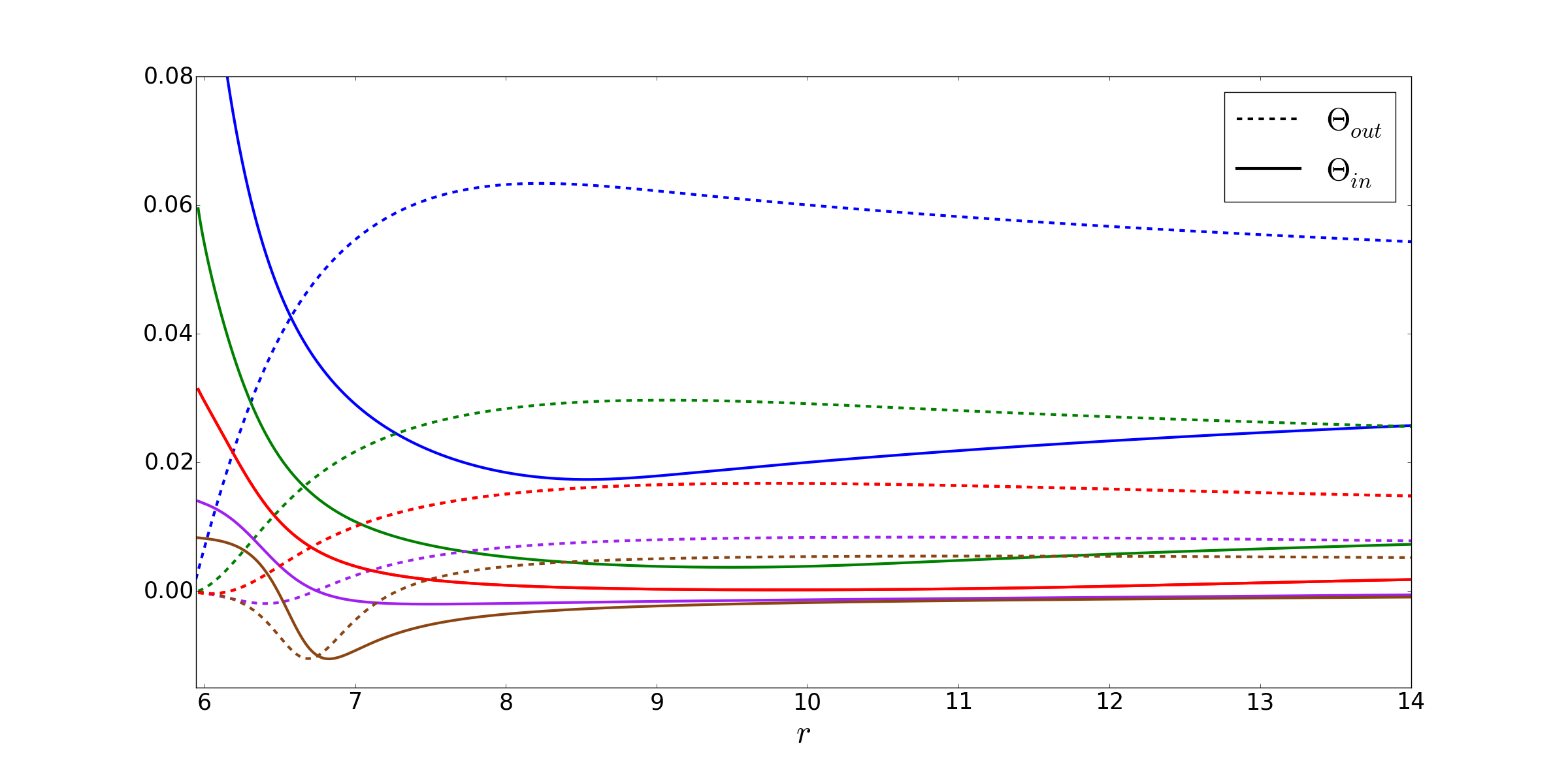}\protect\caption{\label{expansions}The evolution of the expansions outside the wall
in the case of radiation background. The wall is located at $r\sim6$.
Blue curves are the earliest curves in this set. By the definitions
of $\Theta_{in}$ and $\Theta_{out},$ their intersection gives the
position of the wormhole throat.}
\end{figure}

\begin{figure}
\includegraphics[scale=0.3]{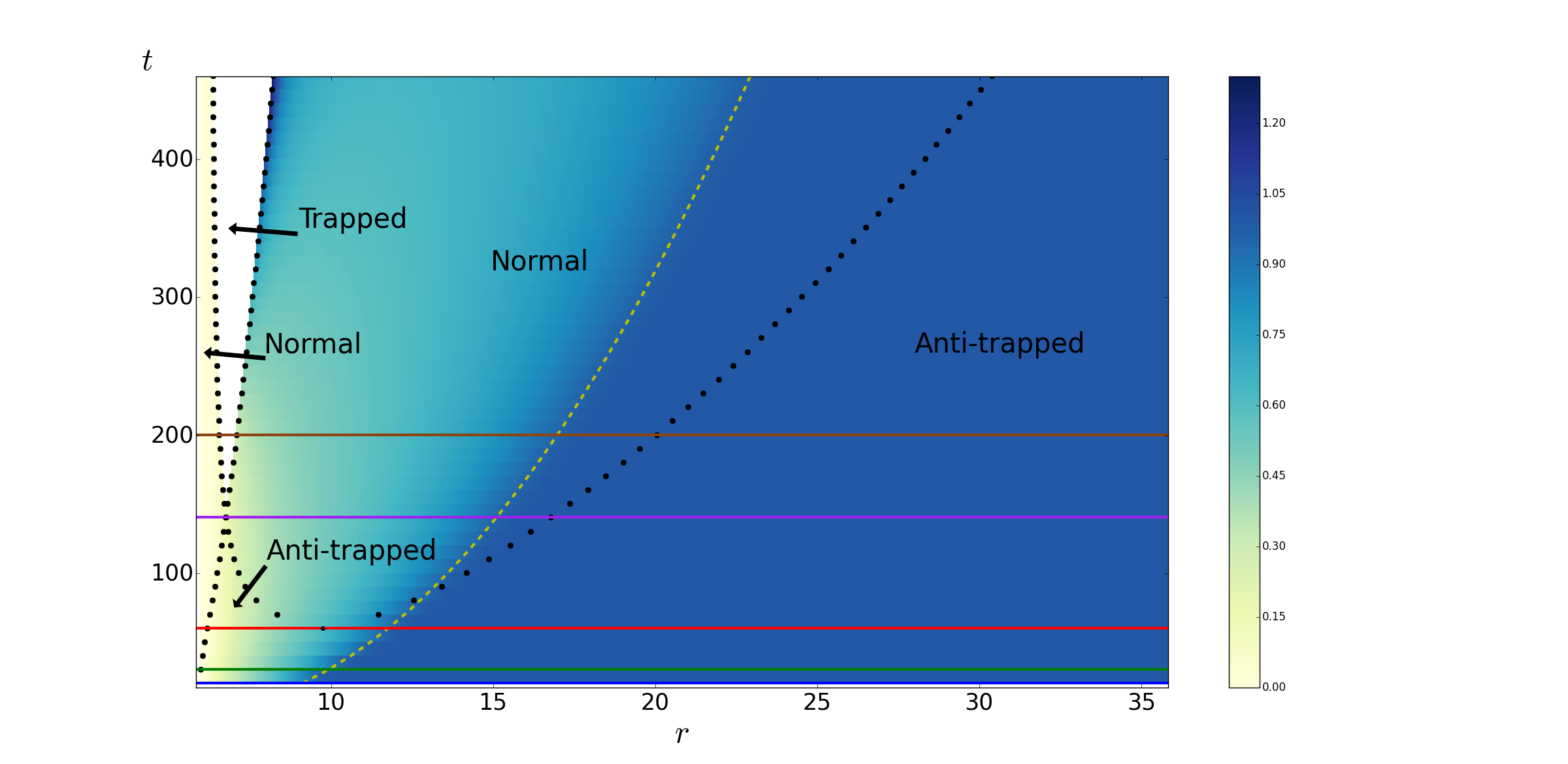}\protect\caption{\label{penrose}Depicted under simulation coordinates, this graph
shows the radiation energy density distribution outside the wall.
The wall is located at $r\sim6$. The energy density has been rescaled
so that $\rho_{FRW}=1$. Two apparent horizons and the cosmological
horizon are shown with black dots. Five constant time slices here
respectively correspond to the five different stages discussed in
text. The yellow dashed curve is the rarefaction wave front trajectory (Eq. (\ref{wavefront})). The intersection of the yellow curve and the black dots indicates the time when the unaffected FRW region first comes within the cosmological horizon.}
\end{figure}

Let us now consider the formation and time evolution of the wormhole.
Fig. \ref{expansions} shows the behavior of the expansion of radial
null congruence as a function of radius, for different times, and
Fig. \ref{penrose} shows the spacetime structure outside the wall,
with the lines separating anti-trapped from normal regions, and normal
regions from trapped regions. These two graphs are obtained from a
simulation of a supercritical wall in radiation background with $r_{i}=6$
and $t_{\sigma}\approx4$. The evolution can be summarized as follows.
(1) Initially, any sphere of radius bigger than the cosmological horizon
$H_{i}^{-1}$ (i.e. with $r>1$) is anti-trapped, so $\Theta_{out},\Theta_{in}>0$
(blue curves and slice). (2) Later, $\Theta_{out}$ reaches zero (green)
at the intersection of the inner white hole horizon and the wall.
Then, the slice begins to cross a normal region inside the throat,
with $\Theta_{out}<0$ and $\Theta_{in}>0$ (note that this is opposite
to what happens in a trivial geometry). An observer in this region
would see himself inside the spherical wall and outside the wormhole.
(3) Afterwards, $\Theta_{in}$ reaches zero (red), at place where
the outer white hole and cosmological horizons emerge in the exterior
region ($r\sim10$ in this example). From that point on, the slice
begins to cross a normal region outside the throat. (4) At a later
time, the marginally anti-trapped surface where $\Theta_{out}=0$
coincides with the surface where $\Theta_{in}=0$ (purple). This corresponds
to a bifurcating marginal surface, and the white hole begins to turn
into a black hole. (5) The slice then leaves the WH region and two
apparent horizons appear with $\Theta_{out}=0$ $(\Theta_{in}<0)$
and $\Theta_{in}=0$ $(\Theta_{out}<0)$ respectively (brown).

\begin{figure}
\includegraphics[scale=0.2]{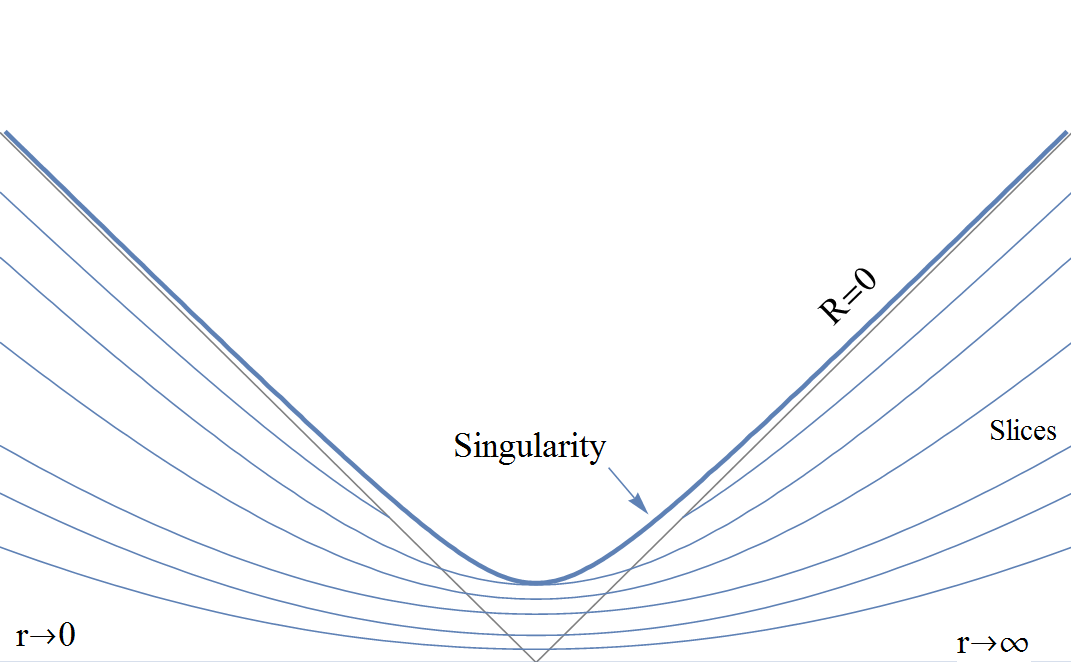}\protect\caption{\label{Singul}This sketch shows how geodesic slicing fails to avoid
the singularity in Schwarzschild spacetime depicted in a Kruskal-Szekeres
diagram \cite{3+1}. We excise the BH region at two apparent horizons
to evolve the spacetime to a much later time.}
\end{figure}

\begin{figure}
\includegraphics[scale=0.2]{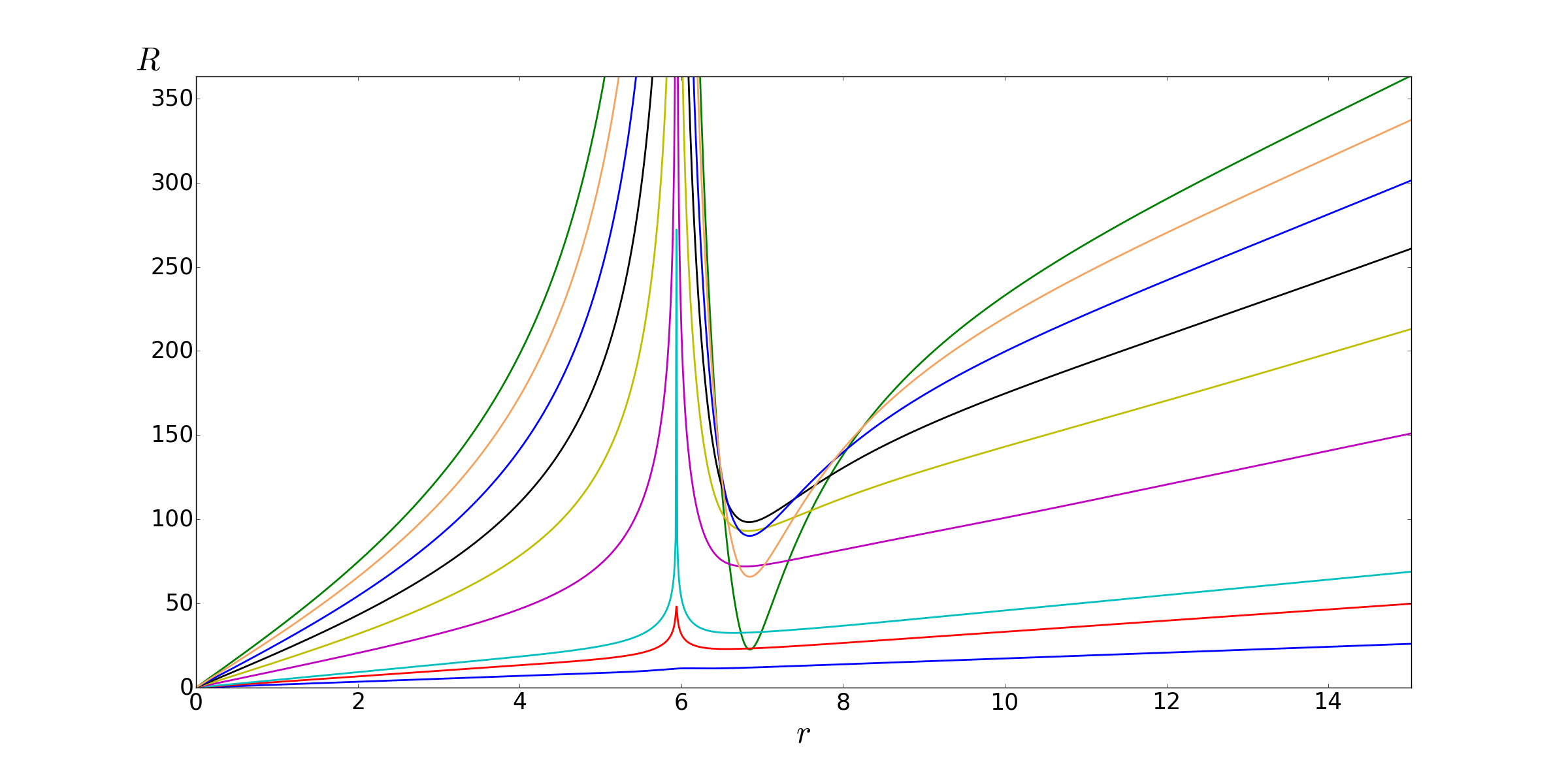}\protect\caption{\label{R-r}The area radius \textbf{$R$} as a function of $r$ for
a supercritical case. The bottom blue curve is the earliest curve
in this set, and the green curve is the latest. Initially $R=r$,
and in the unperturbed exterior region, $R=a(t)r.$ Normally $R$
is monotonic in $r$, but the domain wall distorts the spacetime,
so that a local minimum of $R$ develops with time, which represents
the formation of a wormhole throat. The area radius of the throat
increases at first, then decreases as it collapses to a black hole.
A singularity is encountered when $R$ reaches zero. $R$ on the wall
grows exponentially, which explains the sharp peak at $r\sim6,$ where
the wall is located. }
\end{figure}

\begin{figure}
\includegraphics[scale=0.14]{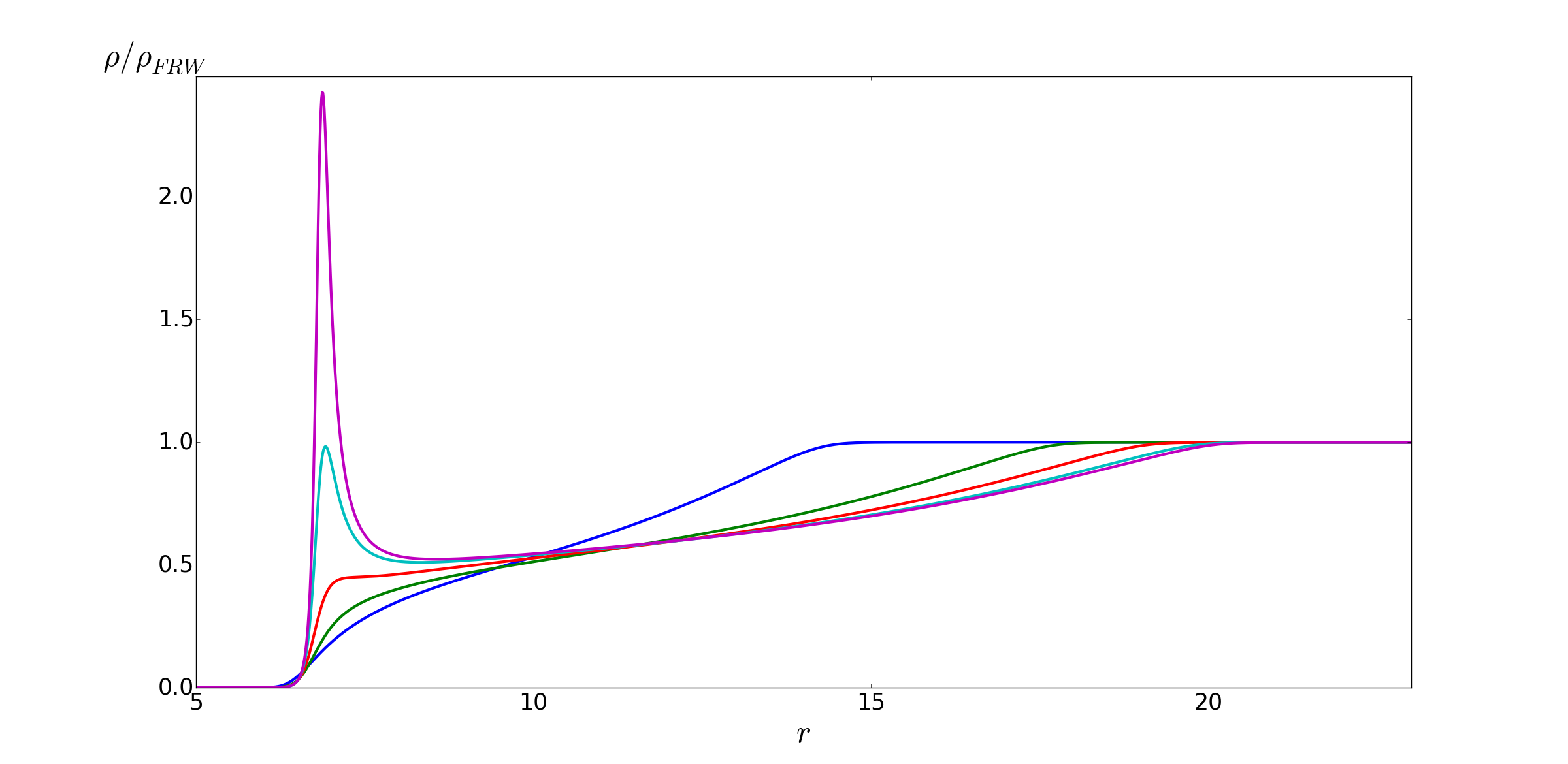}\includegraphics[scale=0.14]{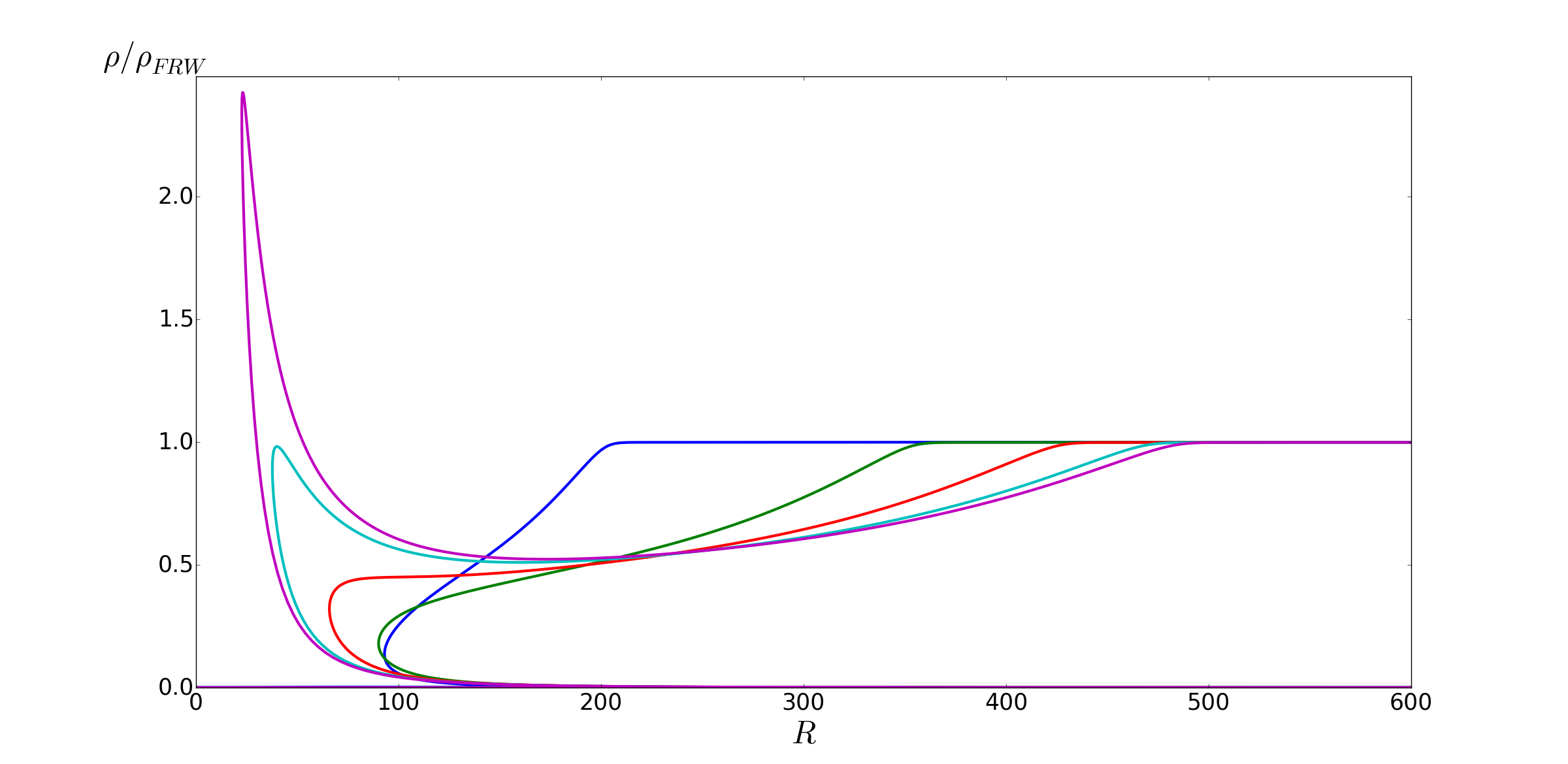}\protect\caption{\label{spike}Evolution of the radiation energy density as the white
hole turns into a black hole. The $x-$axis in the first graph is
the comoving radius $r$, while in the second it is the area radius
$R$. The blue curves are the earliest in this set. After apparent
horizons appear, the area radius of the throat decreases and approaches
zero, and fluid energy density at the throat starts to form a spike
that approaches infinity, which is the singularity. $\rho(R)$ is
double-valued because $R$ goes through the minimum at the throat.
The upper and the lower branches correspond to the radiation energy
density outside and inside the throat, respectively. The wall is located
at $r\sim6.$ $\rho$ in the interior region has decreased significantly
(see Fig. \ref{fireball1}).}
\end{figure}

As in the subcritical case, we can excise the trapped region to avoid
simulation breakdown. Now we need to cut at two apparent horizons
instead of one (see Fig. \ref{Singul}). However, we are still able
to evolve this region for a while to see how the singularity arises.
After the apparent black hole horizons form, the area radius of the
throat begins to decrease and approaches zero (Figs. \ref{R-r} and
\ref{spike}). Correspondingly, the fluid energy density at the throat
starts to increase and eventually goes to infinity. This corresponds
to the spike of growing amplitude in Fig. \ref{spike}.

\subsection{Upper Bound on the Initial BH mass}

At the time $t_{BH}$, two black hole apparent horizons with the same
initial mass $M_{BHi}$ are formed, at the point where $\Theta_{out}$
and $\Theta_{in}$ reach zero in the $(r,t)$ plane. Note that $M_{BHi}$
should have an upper bound $\tilde{M}_{H}$, which is the total mass
enclosed by the horizon crossing radius of the unperturbed FRW region.
Within the cosmological horizon, a spherical surface is normal (i.e.
it's not trapped nor anti-trapped). A contradiction would arise if
the black hole (or white hole) apparent horizon were bigger than the
cosmological horizon, because the region inside the black hole (or
white hole) is trapped (or anti-trapped). Therefore, the Schwarzschild
radius of the black hole cannot exceed the horizon crossing radius
of the unperturbed FRW region.

In the case of dust background, if we neglect the initial wall thickness,
we simply have $\tilde{M}_{H}=M_{H}=({1}/{2})r_{i}^{3},$ because
the FRW boundary outside the wall is unperturbed. In the case of radiation,
however, the perturbation due to the wall propagates outwards as a
rarefaction (or decompression) wave. The speed of the wave is the
speed of sound, $a(dr/dt)={1}/{\sqrt{3}}$, where $a(t)=(t/t_{i})^{1/2}$
and \textbf{$r$ }is the comoving radius of the wave front. Neglecting
the wall thickness, and taking into consideration that $t_{i}=1/2$,
the solution is 
\begin{equation}
r(t)=r_{i}+\frac{1}{\sqrt{3}}\left(\sqrt{2t}-1\right).\label{wavefront}
\end{equation}
The time when this expanding wave front comes within the cosmological
horizon (see Fig. \ref{penrose})
can be found from $a(t)r(t)=2t$, which gives 
\begin{equation}
\tilde{M}_{H}=\tilde{t}_{H}\approx2.8\left(r_{i}-\frac{1}{\sqrt{3}}\right)^{2}\label{tildeM}
\end{equation}
For large $r_{i}$, $\tilde{M}_{H}\approx2.8r_{i}^{2}=5.6M_{H}$. As we shall see, the actual black hole mass is comparable to this upper bound.

\subsection{Inner BH}

\begin{figure}
\includegraphics[scale=0.2]{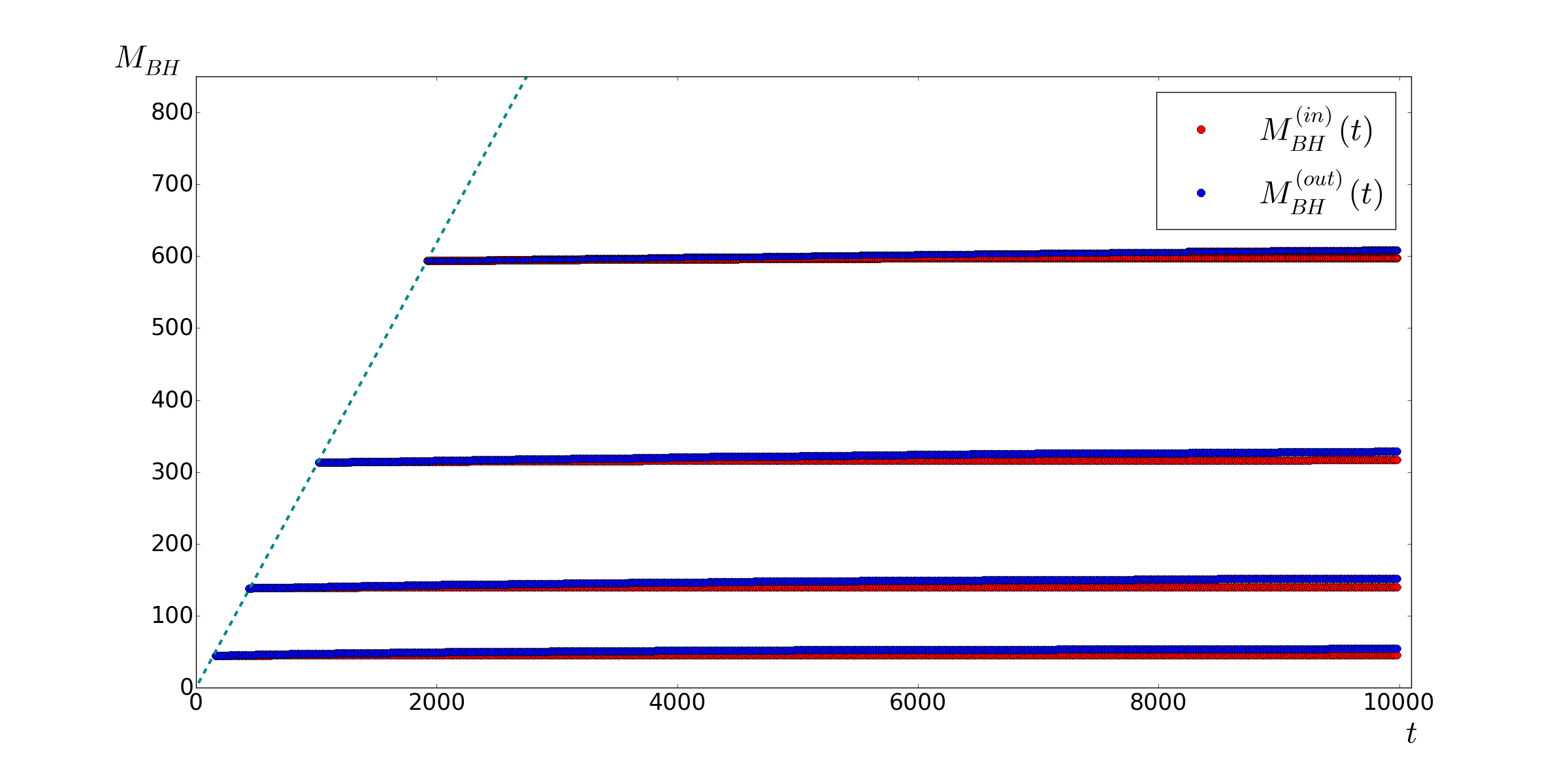}\protect\caption{\label{Md-t}Evolution of $M_{BH}(t)$ for four supercritical walls
in the background of dust with $r_{i}=4,6,8\ \mbox{and }10$ from
the bottom, and $t_{\sigma}\approx3$. When the black holes are formed,
the masses of the inner and the outer black hole are identical. We
find that $M_{BHi}\propto t_{BH}$ (dashed straight line). Then $M_{BH}^{(in)}(t)$
converges very quickly, while $M_{BH}^{(out)}(t)$ grows for a long
time, yet doesn't increase much. The difference between the two is
caused by the wall thickness. If $r_{i}$ is very large, $M_{BHf}^{(in)}\approx M_{BHf}^{(out)}$. }
\end{figure}

\begin{figure}
\includegraphics[scale=0.2]{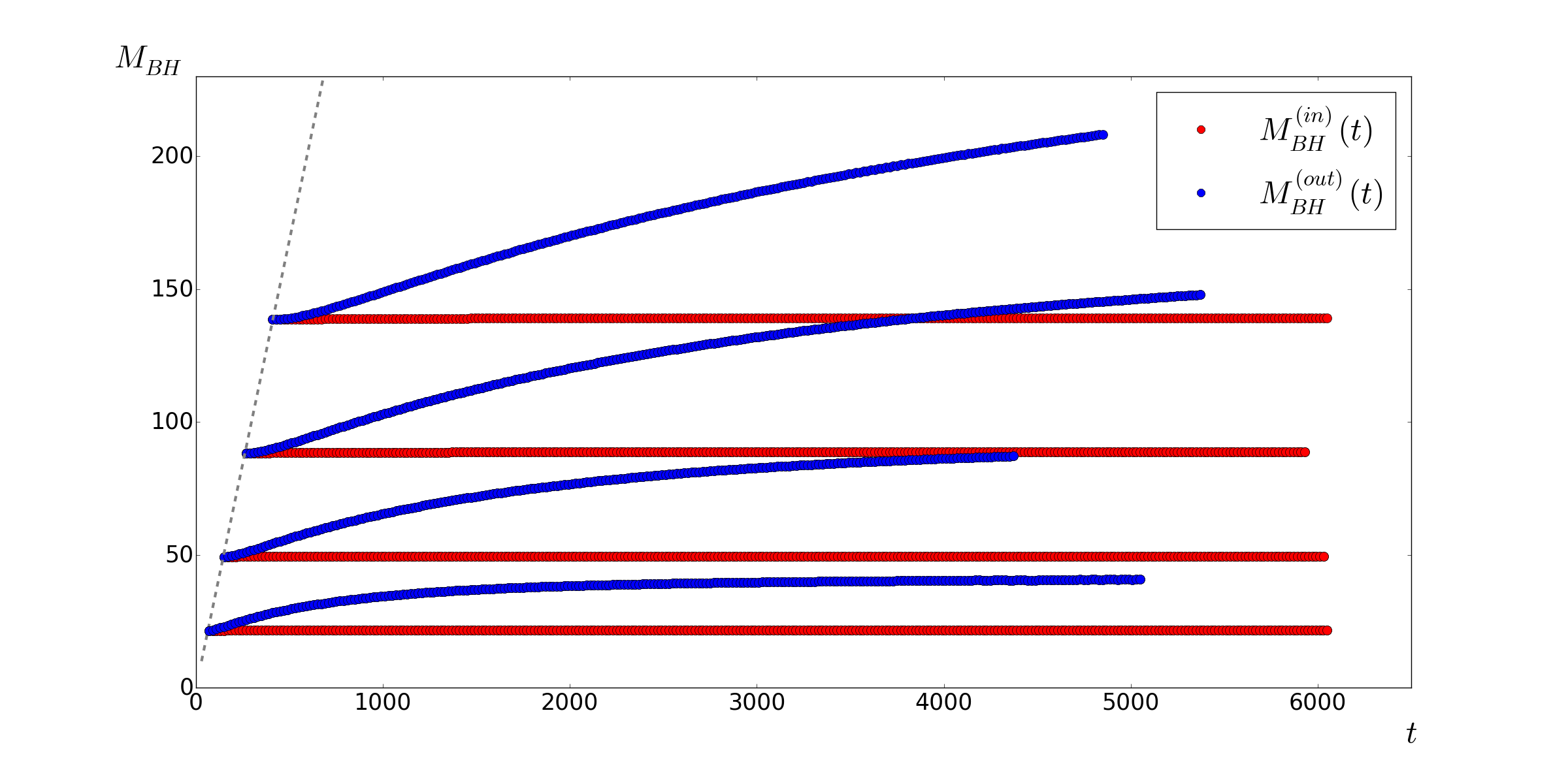}\protect\caption{\label{Mr-t}Evolution of $M_{BH}(t)$ for four supercritical walls
in the background of radiation with $r_{i}=4,6,8\ \mbox{and }10$
from the bottom, and $t_{\sigma}\approx3$. Again we find that $M_{BHi}\propto t_{BH}$
(dashed straight line). $M_{BH}^{(in)}(t)$ converges very quickly,
while $M_{BH}^{(out)}(t)$ grows for a long time, as radiation in
the exterior region keeps falling in. }
\end{figure}

Simulations show that $M_{BHi}$ and $t_{BH}$ are linearly related
(see Figs. \ref{Md-t} and \ref{Mr-t}). In a dust-dominated universe,
$M_{BHi}\approx0.31\ t_{BH}$; for radiation, $M_{BHi}\approx0.34\ t_{BH}.$

\begin{figure}
\includegraphics[scale=0.2]{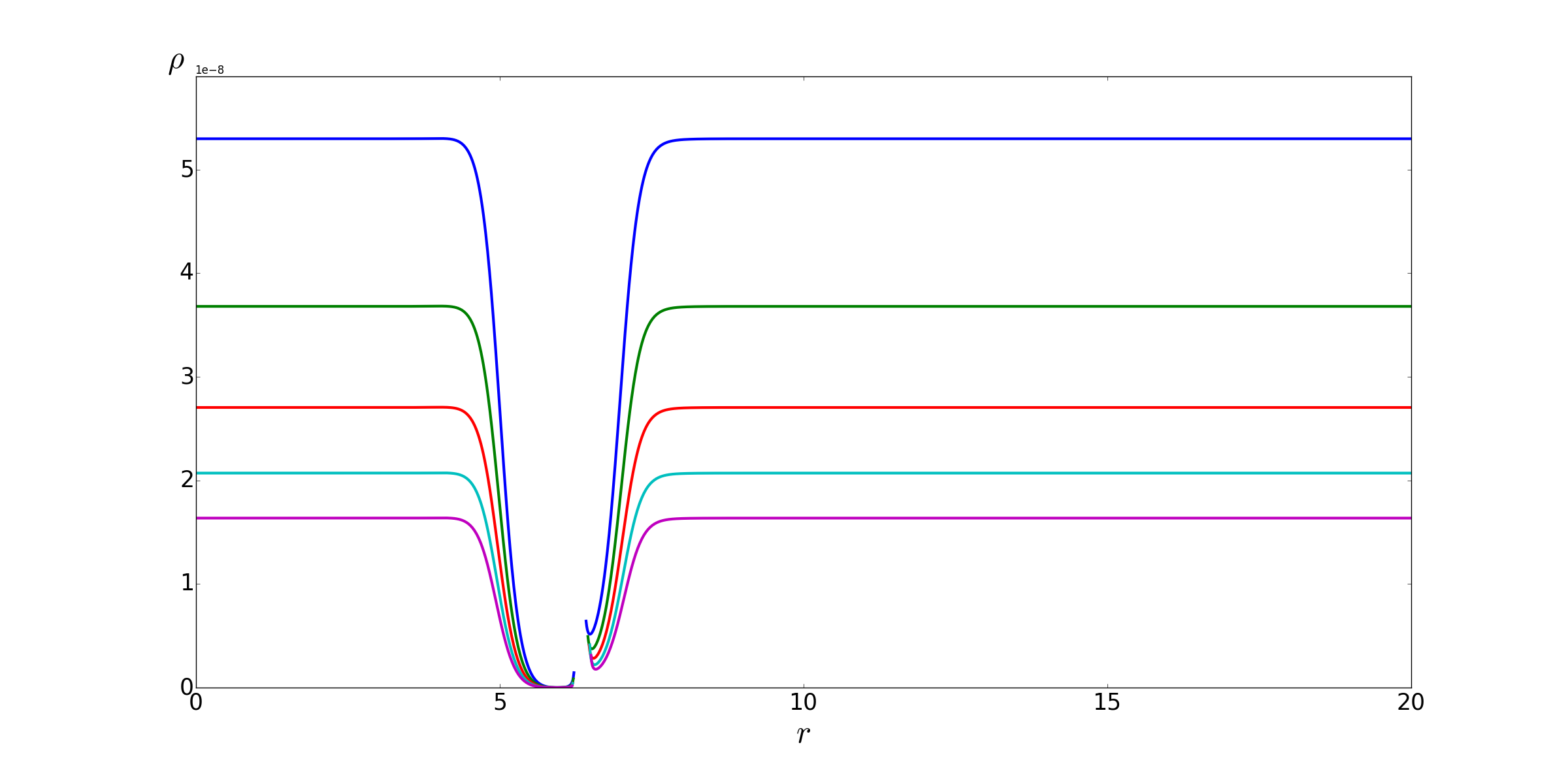}

\includegraphics[scale=0.2]{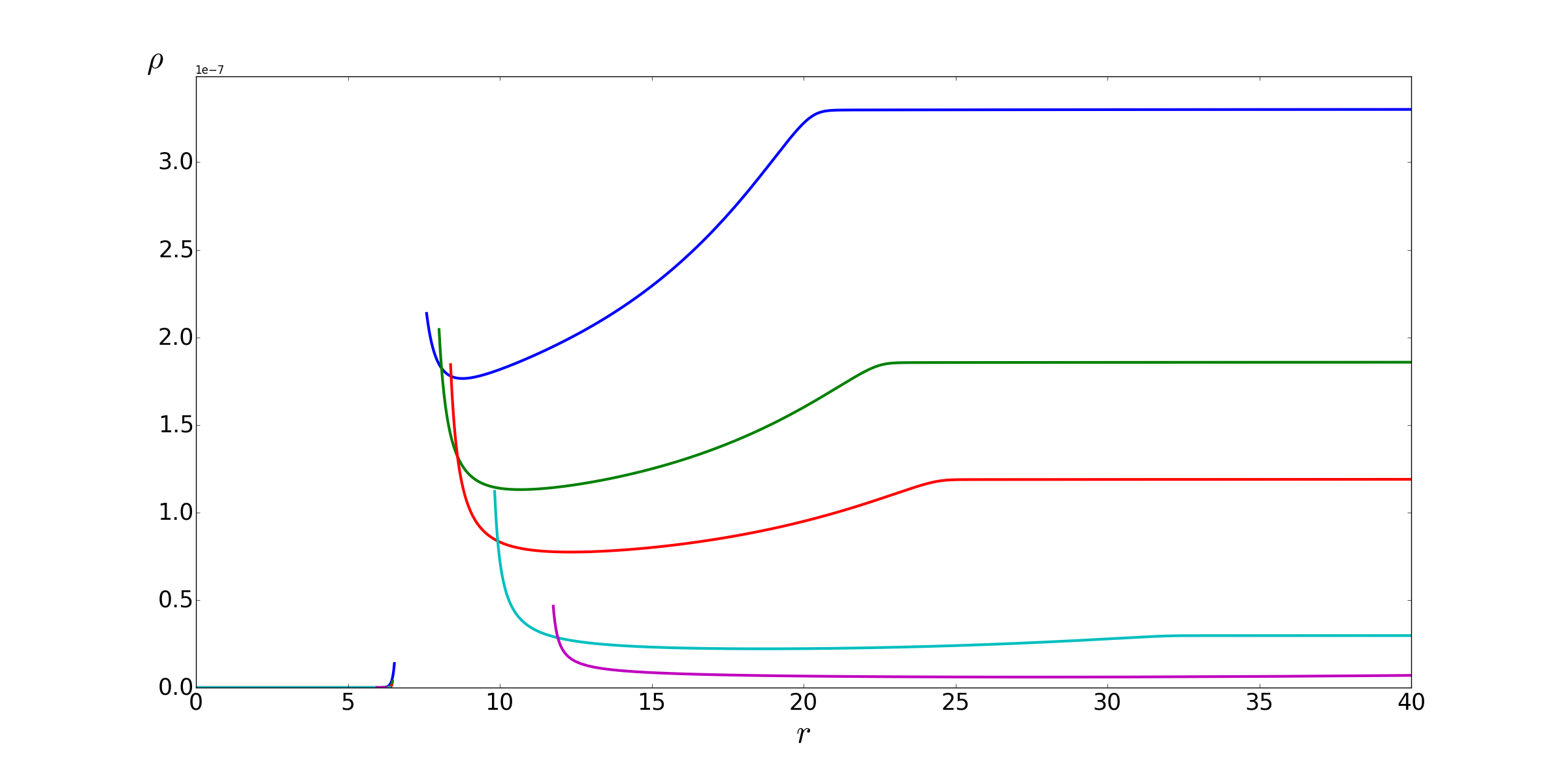}\protect\caption{\label{dust-radiation}The upper and lower graphs illustrate the mass
accretion processes for dust and radiation backgrounds, respectively.
Each graph shows the energy density distributions at four different
moments, with time increasing from top to bottom. The wall is located
at $r\sim6.$ (To avoid simulation breakdown, we excise most of the
spacetime region between the two apparent horizons.) 
Dust between the outer black hole horizon and the exterior FRW boundary
falls in, so $M_{BHf}^{(out)}$ is slightly larger than $M_{BHf}^{(in)}$.
If the wall is very large, this difference is insignificant. In the
case of radiation background, the rarefaction wave propagates outwards,
and the energy density approaches a uniform value. As the energy density
decreases in the exterior FRW region, accretion slows down and the
black hole mass converges to a constant. In radiation case, $\rho$
in the interior region has decreased significantly (see Fig. \ref{fireball1}).}
\end{figure}

After the black holes are formed, fluid nearby falls in and the black
hole masses increase. For the inner black hole the situation is not
very different for dust and radiation because the energy density has
a tiny value between the wall and the apparent horizon (see Fig. \ref{dust-radiation}).
Starting with $M_{BHi},$ $M_{BH}^{(in)}(t)$ converges very quickly
(see Fig. \ref{Md-t} and \ref{Mr-t}), so $M_{BHf}^{(in)}\approx M_{BHi}.$

\begin{figure}
\includegraphics[scale=0.2]{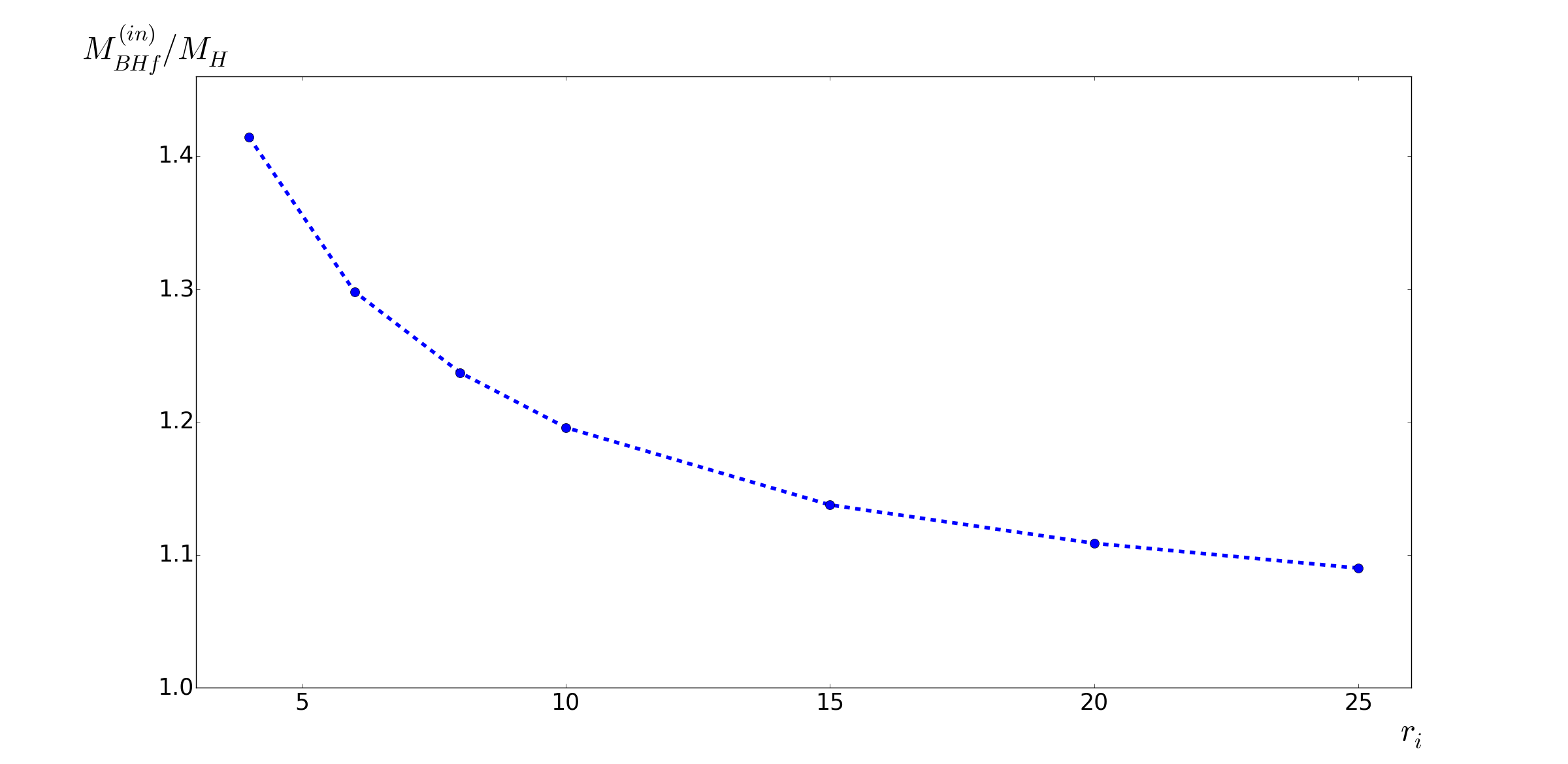}\protect\caption{\label{Md-r}Blue dots are $\frac{M_{BHf}^{(in)}}{M_{H}}$ for seven
supercritical walls in the background of dust with $r_{i}=4,6,8,10,15,20$
and $25$. We can see that the ratio approaches $1$ as $r_{i}$ increases. }
\end{figure}

In the case of dust background, it was argued in Ref. \cite{Garriga}
that for a large supercritical wall, $M_{BHf}^{(in)}\approx M_{H},$
where $M_{H}=(1/2)r_{i}^{3}$ is the enclosed mass within the initial
shell. In Fig. \ref{Md-r} we can see that the ratio ${M_{BHf}^{(in)}}/{M_{H}}$
does approach 1 with an increasing $r_{i}.$ The deviation comes from
the initial wall thickness and energy compensation.

\begin{figure}
\includegraphics[scale=0.2]{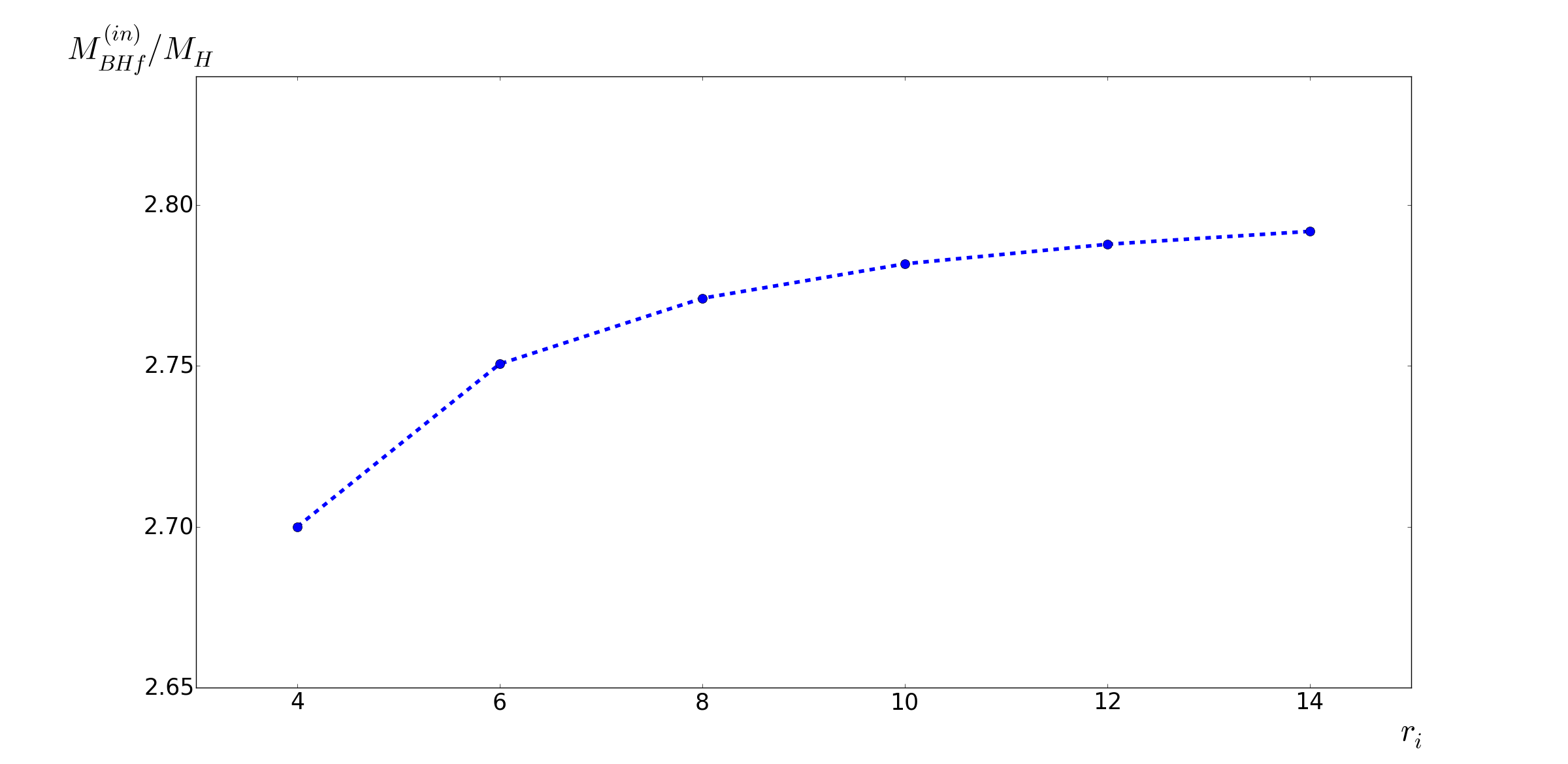}\protect\caption{\label{Mr-r}Blue dots are $\frac{M_{BHf}^{(in)}}{M_{H}}$ for six
supercritical walls in the background of radiation with $r_{i}=4,6,8,10,12\ \mbox{and }14$.
We see that the ratio approaches 2.8 as $r_{i}$ increases. }
\end{figure}

As for the case of radiation, simulations show that ${M_{BHf}^{(in)}}/{M_{H}}\approx2.8$
as $r_{i}$ increases, where $M_{H}=({1}/{2})r_{i}^{2}$ is the enclosed
mass at $t_{H}$ (see Fig. \ref{Mr-r}). This is about half of the
upper bound estimated in Eq. (\ref{tildeM}). As in the case of dust
background, $M_{BHf}^{(in)}$ is independent of the wall's tension.

\begin{figure}
\includegraphics[scale=0.25]{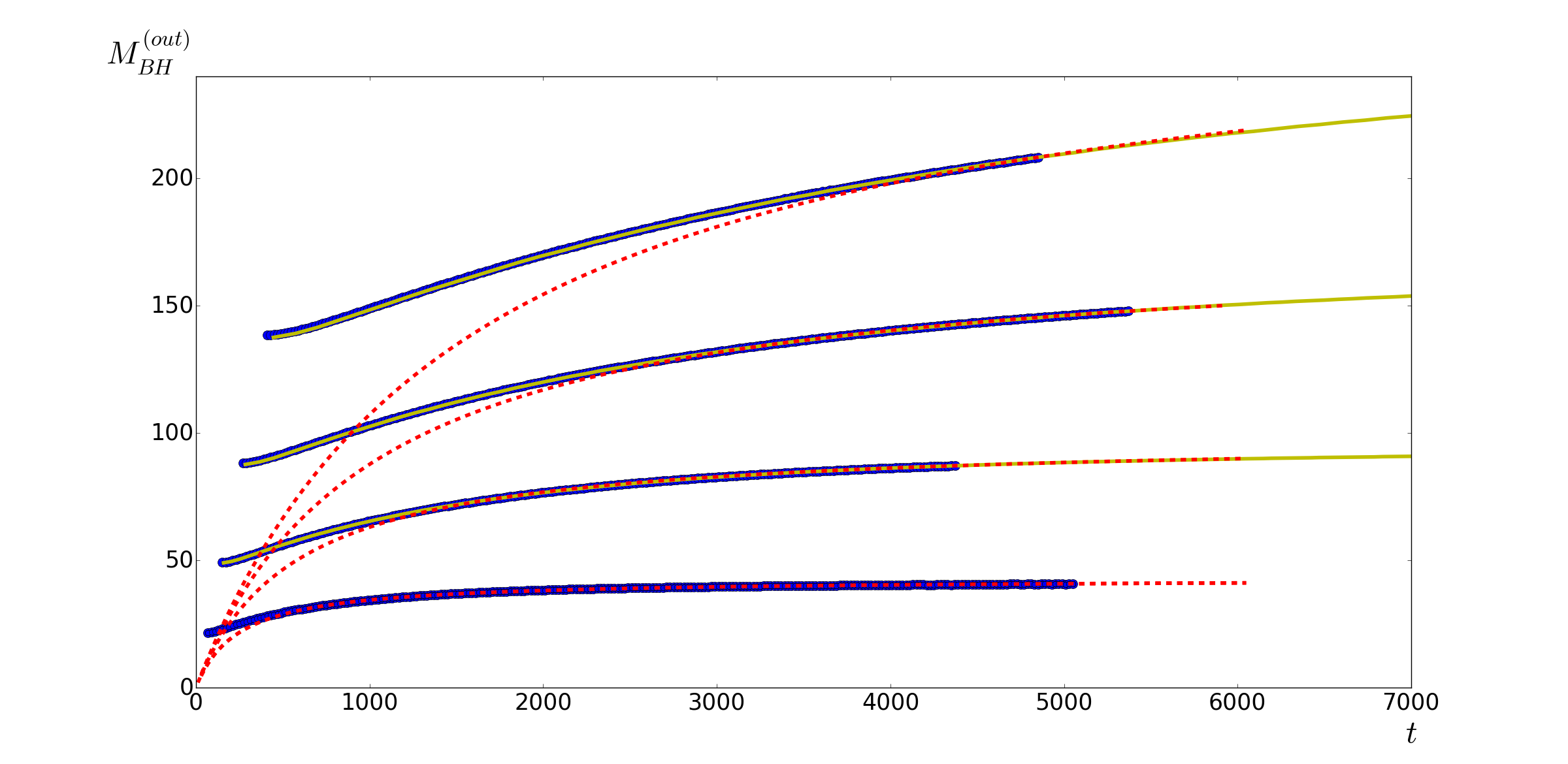}\protect\caption{\label{M-t}This graph shows $M_{BH}^{(out)}$ as a function of time
for four supercritical walls in the background of radiation with $r_{i}=4,6,8\ \mbox{and }10$
respectively (the same walls as those in Fig. \ref{Mr-t}). Blue curves
are from simulations, dashed red curves from Eq. (\ref{accretion}),
and solid yellow curves are from Eq. (\ref{bh}) relative to the wall
with $r_{i}=4$. For $r_{i}=6$, $K=1.512^{2}\approx\left(\frac{6}{4}\right)^{2}$;
For $r_{i}=8$, $K=2.022^{2}\approx\left(\frac{8}{4}\right)^{2}$;
For $r_{i}=10$, $K=2.533^{2}\approx\left(\frac{10}{4}\right)^{2}$. }
\end{figure}

\subsection{Outer BH}

As in the subcritical case, the evolution of $M_{BH}^{(out)}(t)$
is very different for different fluids (Fig. \ref{dust-radiation}).
Dust between the apparent horizon and the edge of the exterior FRW
region eventually falls into the black hole, while the FRW region
is unaffected. If $r_{i}$ is sufficiently large, such mass accretion
is negligible, and we have $M_{BHf}^{(out)}\approx M_{H}$ (see Fig.
\ref{Md-t}).

However, in the case of radiation, since fluid keeps falling in, $M_{BH}^{(out)}(t)$
grows for a long time before it converges. We find that Eq. (\ref{accretion})
applies here as well at later times, again with $F\approx3.8$. Moreover,
it turns out that $M_{BH}^{(out)}(r_{i},t)$ satisfies a scaling relation 
(see Fig. \ref{M-t}). For two walls labeled by 1 and 2 respectively,
we find that 
\begin{equation}
M_{BH2}^{(out)}(Kt)=KM_{BH1}^{(out)}(t)\label{bh}
\end{equation}
where 
\begin{equation}
K\approx\frac{M_{BHi2}}{M_{BHi1}}\approx\left(\frac{r_{i2}}{r_{i1}}\right)^{2}
\end{equation}
Since $M_{BH}^{(out)}(t)\to M_{BHf}^{(out)}$ as $t\to\infty$, it
follows that 
\begin{equation}
\frac{M_{BHf2}^{(out)}}{M_{BHf1}^{(out)}}\approx\left(\frac{r_{i2}}{r_{i1}}\right)^{2}.
\end{equation}
Therefore the relation between $M_{BHf}^{(out)}$ and $M_{BHi}$ (or
$M_{H}$) can be determined by looking at a single case, say $r_{i}=10.$
By reading $M_{0}$ and $t_{0}$ from simulations at a sufficiently
late time, Eq. (\ref{accretion}) gives $M_{BHf}^{(out)}\approx276\approx5.5M_{H}\approx2M_{BHi}$.
Hence we conclude that for supercritical domain walls in the background
of radiation, the mass of the outer black hole roughly doubles after
its formation.

\section{Fireball}

\label{fireball}

\begin{figure}
\includegraphics[scale=0.2]{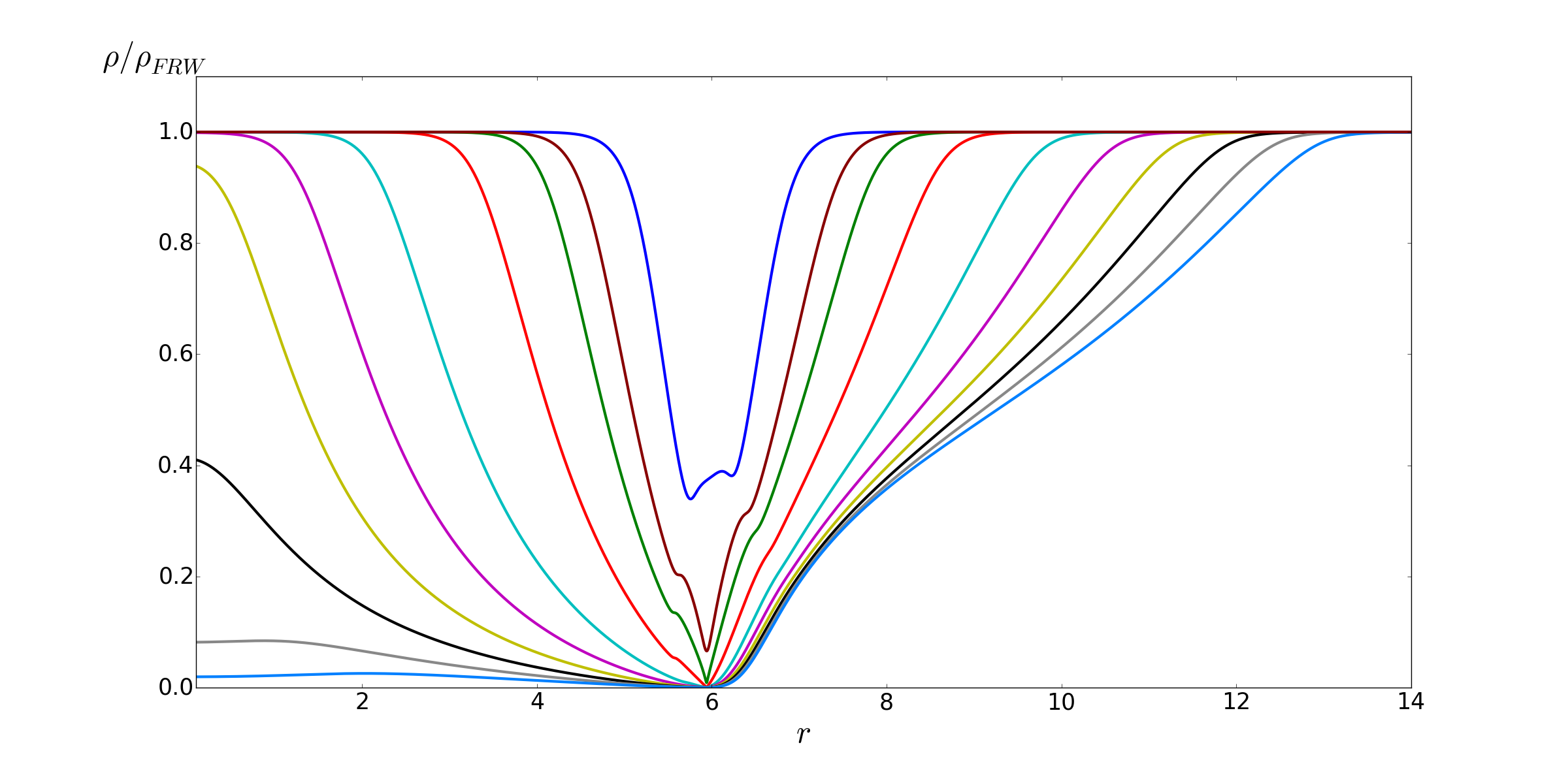}\protect\caption{\label{fireball1}Propagation of the rarefaction waves. As the wave
front reaches the center, the radiation energy density begins to drop
much faster than that in the exterior region.}
\end{figure}

\begin{figure}
\includegraphics[scale=0.2]{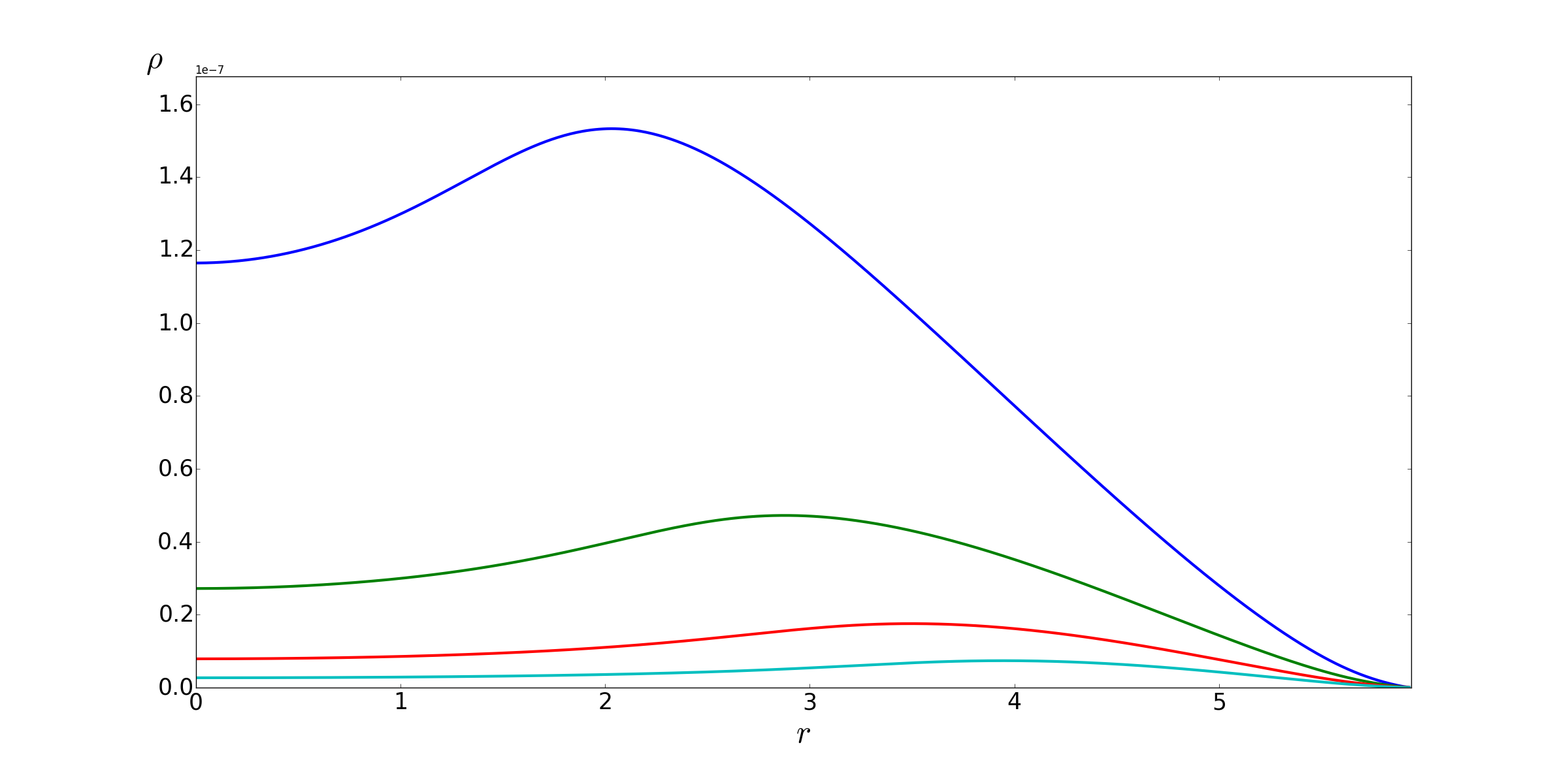}\protect\caption{\label{fireball2}As the wave front reaches the center, a reflected
wave is produced and begins to propagate outwards.}
\end{figure}

\begin{figure}
\includegraphics[scale=0.2]{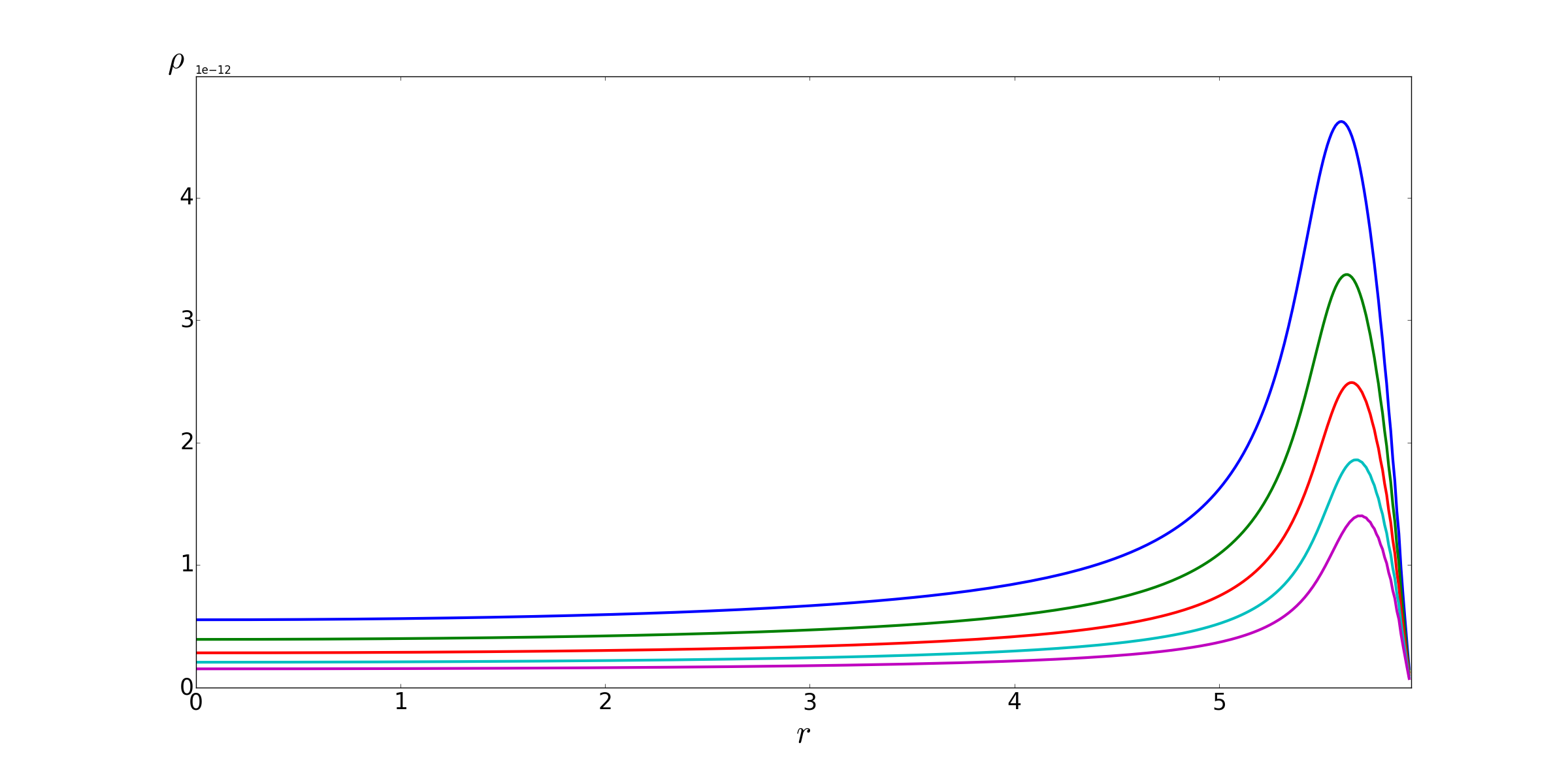}\protect\caption{\label{firebal3}As the reflected wave moves outwards, a shell is
formed where most energy is concentrated and moves nearly at the speed
of light, leaving an empty region inside the fireball. }
\end{figure}

\begin{figure}
\includegraphics[scale=0.2]{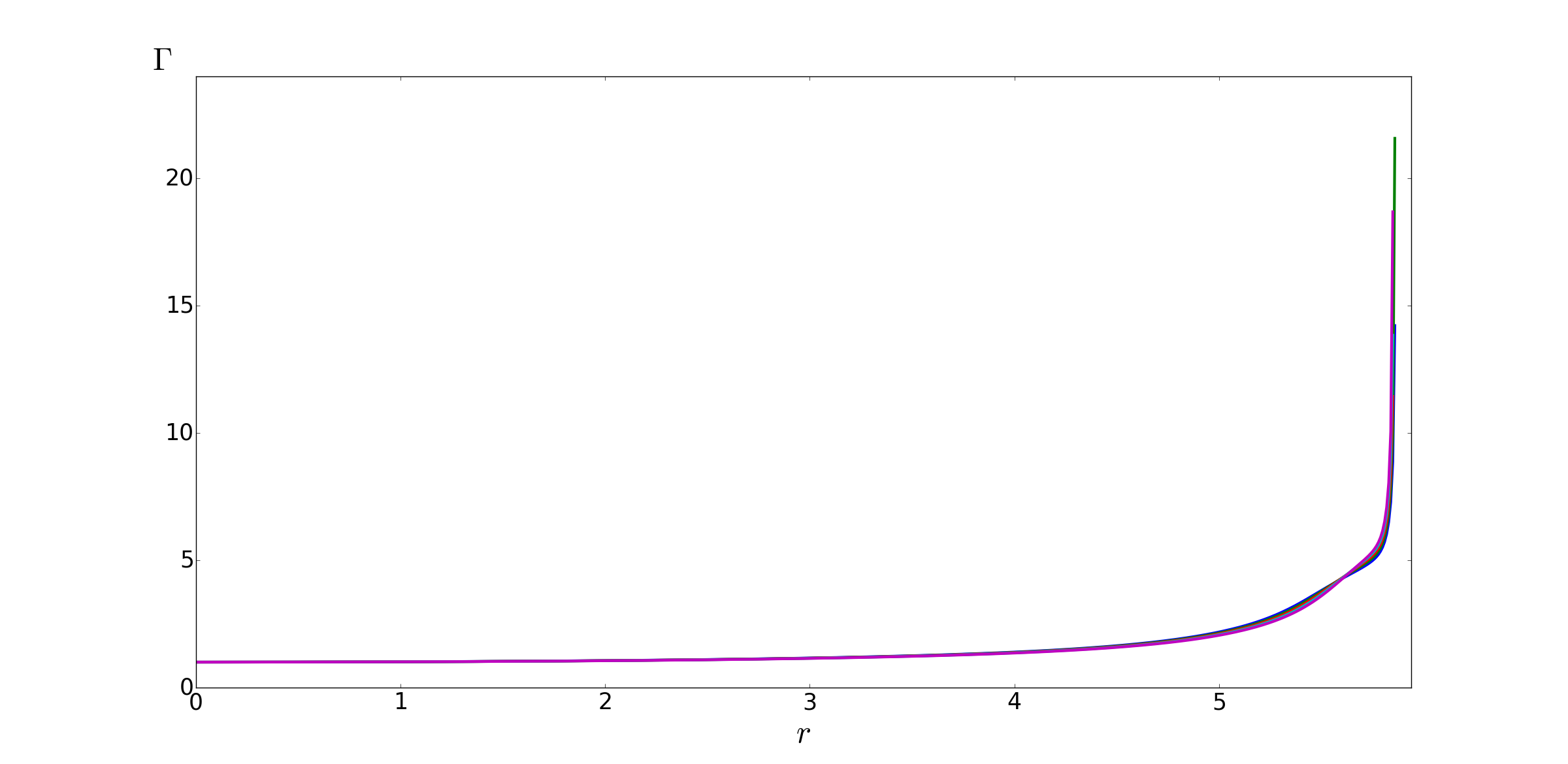}\protect\caption{\label{fireball4}The Lorentz factor $\Gamma=\frac{1}{\sqrt{1-v^{2}}}$
as a function of $r$ as the shells expand in Fig. \ref{firebal3}. }
\end{figure}

For completeness, here we consider the evolution of the interior region
of a supercritical wall in radiation background. Once the empty layers
are formed, as the wall inflates away, we are left with a relativistic
fireball expanding into vacuum. The dynamics of this system is of
interest in itself, because similar fireballs may be at the origin
of $\gamma$-ray bursts (GRBs). Analytical and numerical investigations
of the fireball expansion as applied to the GRB events have been done
in numerous works \cite{Goodman,Paczynski,Piran,Bisnovatyi,Kobayashi}.

Let us briefly summarize our numerical results in the present context.
In the case of radiation background, a rarefaction wave is produced
near the wall and then propagates towards the center, while the energy
density near the center keeps decreasing as in an FRW universe. If
the initial wall thickness can be neglected, the motion of the wave
front is given by 
\begin{equation}
r(t)=r_{i}-\frac{1}{\sqrt{3}}\left(\sqrt{2t}-1\right),
\end{equation}
where we have used that, in our units, $t_{i}=1/2$. When the wave
reaches the center, at time 
\begin{equation}
t=\frac{1}{2}\left(\sqrt{3}r_{i}+1\right)^{2},
\end{equation}
the energy density near the center begins to drop much faster than
that in the exterior FRW region (see Fig. \ref{fireball1}).  The reason is that there is no external pressure on the fireball, so it freely expands into the empty space which is opened up by the receding wall.  Numerically we find $\rho(r=0,t) \propto t^{-n}$ with $n\approx 8$.  Once the rarefaction wave hits the center, a reflected wave moves outwards (see Fig. \ref{fireball2}). Then a leading shell is formed where most energy is concentrated and moves nearly at the speed of light (see Figs. \ref{firebal3} and \ref{fireball4}), leaving an empty region inside the fireball.  This evolution scenario is in a qualitative agreement with earlier studies of relativistic fireballs  \cite{Goodman,Paczynski,Piran,Bisnovatyi,Kobayashi}.

\section{Summary and Conclusions}

In this paper we have numerically studied the gravitational collapse
of spherical domain walls embedded in an FRW universe. The resulting
spacetime structure depends on whether the acceleration time-scale
$t_{\sigma}=(2\pi G\sigma)^{-1}$ associated with the gravitational
field of the domain wall is larger or smaller than $t_{H}(r_{i})$.
Here, $r_{i}$ is the initial radius of the wall at the time when
inflation ends, and $t_{H}$ is the time at which such comoving scale
would cross the horizon in an undisturbed FRW universe.

Subcritical walls are those with $t_{\sigma}\gg t_{H}$. Their gravitational
field can be neglected at least up to the time $t_{H}$. At that point,
they start shrinking under the pull of their own tension and eventually
collapse, forming ordinary black holes. We developed a semi-analytical
method to obtain the trajectory of subcritical walls, and the initial
mass of the resulting black holes. For walls that collapse in a dust-dominated
universe, we find that the initial mass is given by $M_{BHi}\approx5\sigma t_{H}^{2}$,
while for walls collapsing in a radiation-dominated universe we find
$M_{BHi}\approx30\sigma t_{H}^{2}$. These results are in good agreement
with the numerical studies of Refs. \cite{Tanahashi,Garriga} for
the case of a matter dominated universe, and Ref. \cite{Garriga}
for the radiation-dominated case. We then turned to simulations to
study the subsequent accretion of nearby fluid into the black hole.

For the matter dominated case, we find that the final mass is given
by $M_{BHf}\approx(3/4)t_{H}/G$. This result is in agreement with
Ref. \cite{Tanahashi}, and the situation can be described as follows.
In our model, matter can freely pass through the domain wall, interacting
with it only gravitationally. Nonetheless, all of the matter that
was originally contained within the wall will either remain within
the wall until the black hole forms, or it will exit the wall with
less than the escape velocity, eventually falling back into the black
hole.

On the other hand, for subcritical walls in the radiation-dominated
universe, mass accretion turns out to be more modest. It turns out
that this is well described by Zeldovich and Novikov's model, Eq.
(\ref{Zeldovich}), with an accretion efficiency $F\approx3.8$, and
we find that $M_{BHf}\approx\alpha M_{BHi}$ where $1<\alpha<2$,
and $\alpha\approx2$ as we approach the critical limit where $t_{\sigma}\sim t_{H}$.

Supercritical walls are those with $t_{\sigma}\ll t_{H}$. At the
time $t_{\sigma}$ a supercritical wall starts pushing matter away
from it, while growing in size at an exponential rate. To accommodate
such behavior, the wall moves through a wormhole into a baby universe,
where its exponential expansion continues forever. The wormhole pinches
off on a time-scale of order $t_{H}$, leaving behind two black hole
horizons at the place where the wormhole mouths used to be. One of
the black hole horizons faces the baby universe, while the other one
faces the exterior FRW universe. At the time of formation, the interior
and exterior apparent horizons coincide.

For the case of dust, the initial radius of such apparent horizons
is given by $2GM_{BHi}^{(in)}=2GM_{BHi}^{(out)}\approx(3/2)t_{H}$.
Note that this coincides with the Hubble radius at the time $t_{H}$.
Supercritical walls do not accrete any substantial amount of matter.
All of the matter that was initially inside the wall remains inside
the wall, and never gets close to the black hole horizon which is
facing the baby universe. Matter which is initially outside the wall
follows the flat FRW geodesic motion, with the escape velocity relative
to the black hole, so it does not accrete either. The large supercritical
black holes form as a consequence of the wormhole evolution, without
any matter ever crossing the black hole horizons \cite{Garriga}.

For the case of radiation, simulations show that the initial size
of the apparent horizons is given by $2GM_{BHi}^{(in)}=2GM_{BHi}^{(out)}\approx5.6t_{H}\approx\tilde{t}_{H}$.
Here, $\tilde{t}_{H}\approx3(\sqrt{3}-1)^{-2}t_{H}$ is defined as
the time when the region affected by the wall motion comes within
the Hubble radius. In this expression for $\tilde{t}_{H}$, we have
taken into consideration that the affected comoving region grows in
time, since as soon as the wall starts its exponential expansion at
the time $t_{\sigma}\ll t_{H}$, a rarefaction wave propagates outward
into the FRW region at the speed of sound. For instance, in the simulation
represented in Fig. \ref{penrose}, we have $\tilde{t}_{H}\approx100H_{i}^{-1}$.
This corresponds to the time when the dotted line representing the
cosmological horizon intersects the boundary of the uniform region
unaffected by collapse. Note that the initial size of the apparent
horizon is approximately half the size of the cosmological horizon
at the time $\tilde{t}_{H}$. On the other hand, $\tilde{t}_{H}$
precedes the time $t_{BH}$ when the black hole apparent horizons
form, which is represented in Fig. \ref{penrose} by a purple line.
Hence, the initial size of the black holes is smaller than half of
the cosmological horizon size at the time $t_{BH}$. 

Mass accretion does not affect the size of the inner apparent horizon
$M_{BH}^{(in)}(t)$ much, because fluid near the inner black hole
has been pushed away by the wall. On the other hand, mass accretion
onto the outer black hole horizon proceeds according to Eq. (\ref{Zeldovich}),
with the same accretion efficiency factor as in the case of subcritical
walls. Furthermore, we find that $M_{BH}^{(out)}(r_{i};t)$ satisfies a scaling relation, so it is enough to consider a single value
of the large supercritical initial wall radius $r_{i}$. From the
asymptotic behavior of this solution at large times, we find that
the mass grows approximately by a factor of 2, so that the final Schwarzschild
radius is approximately given by the Hubble radius at the time $\tilde{t}_{H}$,
that is, $2GM_{BHf}^{(out)}\approx2\tilde{t}_{H}\equiv\tilde{H}^{-1}$.

\begin{table}
\begin{tabular}{|c|c|c|}
\hline 
BH Mass  & Dust  & Radiation\tabularnewline
\hline 
\hline 
$M_{BHi}$  & $\approx5\sigma t_{H}^{2}$  & $\approx30\sigma t_{H}^{2}$\tabularnewline
\hline 
$M_{BHf}$  & $3t_{H}/4G$  & $(1\text{-}2)M_{BHi}$\tabularnewline
\hline 
\end{tabular}

\protect\caption{\label{table1}BH masses in subcritical case}
\end{table}



\begin{table}
\begin{tabular}{|c|c|c|}
\hline 
BH Mass  & Dust  & Radiation\tabularnewline
\hline 
\hline 
$M_{BHi}^{(in)}$  & $3t_{H}/4G$  & $2.8t_{H}/G$\tabularnewline
\hline 
$M_{BHf}^{(in)}$  & $3t_{H}/4G$  & $2.8t_{H}/G$\tabularnewline
\hline 
$M_{BHi}^{(out)}$  & $3t_{H}/4G$  & $2.8t_{H}/G$\tabularnewline
\hline 
$M_{BHf}^{(out)}$  & $3t_{H}/4G$  & $5.6t_{H}/G$\tabularnewline
\hline 
\end{tabular}

\protect\caption{\label{table2}BH masses in supercritical case}
\end{table}

By continuity, there should be a critical state in which the wall
retains a constant radius. In fact, as the critical regime is approached
from the subcritical case, a wormhole throat would be formed outside
the wall and a baby universe created. If the wall gravity is still
not sufficiently strong, the wall would eventually collapse, and the
baby universe would disappear. However, if the wall is balanced by
its tension and repulsive gravity, it would retain an asymptotically
constant size. This is an unstable state, and the wall would either
collapse or grow forever in the baby universe under a small perturbation.

Black hole masses in different scenarios are summarized in Table \ref{table1}
and Table \ref{table2}. 
Our results agree with nearly all the expectations in Ref. \cite{Garriga},
except that the supercritical black hole mass in radiation background
is about 5.6 times larger than the upper bound estimated there. This
corresponds to the ratio between $\tilde{t}_{H}$ and $t_{H}$. Aside
from this overall factor, the black hole mass distribution obtained
in Ref. \cite{Garriga} remains unchanged.

Although the present work is motivated by the scenario where domain
walls nucleate by quantum tunneling during inflation, some of our
results may also be useful in scenarios where the sites of domain
walls after inflation are determined by quantum diffusion of light
fields during inflation \cite{Rubin,Khlopov}. This connection is
worth investigating, and is left for further research. We also note
that our analysis of mass accretion onto black holes may be applicable
more widely, to models where primordial black holes are not formed
by domain walls, but by other mechanisms (see, e.g., \cite{Carr3,Khlopov2,Frampton,Juan}). 
\begin{acknowledgments}
We would like to thank Jose Juan Blanco-Pillado and Xiaozhe Hu for
helpful discussions on some numerical simulation issues. This work
was supported in part by the National Science Foundation under grant
PHY-1518742 (H.D. and A.V.) and by MEC FPA2013-46570-C2-2-P, FPA2016-76005-C2-2-P,
MDM-2014-0369 of ICCUB (Unidad de Excelencia ``Mar\'ia de Maeztu\textquotedblright ),
AGAUR 2014-SGR-1474 (J.G.). 
\end{acknowledgments}

\section*{Appendix}

In this Appendix we use Israel's junction conditions to derive an
approximate equation of motion for a subcritical domain wall. In the
subcritical case, when the gravitational effect of the wall is not
significant, we may assume that the interior region is a flat FRW
universe with metric 
\begin{equation}
ds^{2}=-dt^{2}+a^{2}(t)(dr^{2}+r^{2}d\Omega^{2}).\label{FRW}
\end{equation}
In the exterior region the metric can be expressed as 
\begin{equation}
ds^{2}=-dt^{2}+a_{1}(r,t)^{2}dr^{2}+a_{2}(r,t)^{2}r^{2}d\Omega^{2}.\label{metric2}
\end{equation}
For a thin wall, we can use the junction conditions to match metrics
(\ref{FRW}) and (\ref{metric2}) with $a_{1}=a_{2}=a$ inside the
wall. We closely follow the method and notations in \cite{Tanahashi}.

Let the trajectory of the wall be $(t(\tau),r(\tau),0,0),$ where
$\tau$ is the proper time on the wall, then the tangent vector is
$v^{\mu}=(t_{,\tau},r_{,\tau},0,0)$, where $_{,\tau}\equiv d/d\tau.$
Assuming $\dot{t}$ to be positive, $v^{\mu}v_{\mu}=-1$ gives $t_{,\tau}=\sqrt{1+a_{1}^{2}r_{,\tau}^{2}}.$
(There's an abuse of symbols here since $t$ and $r$ are used as
coordinates as well as the trajectory).

Let $\xi^{\mu}$ be the normal unit vector on the wall; then $\xi^{\mu}v_{\mu}=0$
and $\xi^{\mu}\xi_{\mu}=1$ give $\xi^{\mu}=(a_{1}r_{,\tau},t_{,\tau}/a_{1},0,0)$
and $\xi_{\mu}=(-a_{1}r_{,\tau},a_{1}t_{,\tau},0,0)$. The induced
metric on the wall is $h_{\mu\nu}=g_{\mu\nu}-\xi_{\mu}\xi_{\nu}$
and the extrinsic curvature is $K_{\mu\nu}=h_{\mu}^{\ \alpha}\nabla_{\alpha}\xi_{\nu}$,
where $\nabla$ is the covariant derivative operator for 4-spacetime.

We define the brackets $[Q]_{0}\equiv Q_{out}-Q_{in}$, and $\{Q\}_{0}\equiv Q_{out}+Q_{in}$,
where ``out'' and ``in'' denote matching quantities on different
sides of the hypersurface. Also we use the notation $\bar{Q}\equiv(Q_{out}+Q_{in})/2.$
Then the first junction condition is $[h_{\mu\nu}]_{0}=0$, and the
second junction condition is $[K_{\mu\nu}]_{0}=8\pi(-S_{\mu\nu}+Sh_{\mu\nu}/2)$,
where $S_{\mu\nu}=-\sigma h_{\mu\nu}$ is the energy-momentum tensor
of the domain wall. The equation of motion for the wall is $S_{\mu\nu}\bar{K}^{\mu\nu}=[T_{\mu\nu}\xi^{\mu}\xi^{\nu}]_{0}.$
Lastly, energy conservation can be expressed as $h_{\mu}^{\ \alpha}\nabla_{\alpha}S_{\ \nu}^{\mu}=-[T_{\mu\alpha}\xi^{\mu}h_{\ \nu}^{\alpha}]_{0}$.


The $(t,t)$ and $(\theta,\theta)$ components of the first junction
condition give 
\begin{equation}
a_{1}^{2}=a_{2}^{2}=a^{2}.
\end{equation}
Therefore $a_{1,\tau}=a_{2,\tau}=a_{,\tau}$, which gives 
\begin{equation}
\dot{a}t_{,\tau}=a_{,\tau}=t_{,\tau}\dot{a_{1}}+r_{,\tau}a_{1}^{\prime}=t_{,\tau}\dot{a_{2}}+r_{,\tau}a_{2}^{\prime}.
\end{equation}
Furthermore, the $(\theta,\theta)$ component of the second junction
condition gives 
\begin{equation}
\left[\xi^{\mu}\partial_{\mu}\ln(a_{2}r)\right]_{0}=-4\pi\sigma,\label{junction1}
\end{equation}
and the $(\tau,\tau)$ component gives 
\begin{equation}
\left[\xi_{\mu}D_{\tau}v^{\mu}\right]_{0}=-4\pi\sigma,\label{junction2}
\end{equation}
where $D_{\tau}v^{\mu}=\partial_{\tau}v^{\mu}+\Gamma_{\lambda\sigma}^{\mu}v^{\lambda}v^{\sigma},$
with $\Gamma_{\lambda\sigma}^{\mu}$ the Chistoffel symbols in 4-spacetime.

A perfect fluid has energy-momentum tensor $T_{\mu\nu}=(\rho+p)u_{\mu}u_{\nu}+pg_{\mu\nu}$.
Then the equation of motion is 
\begin{equation}
\left\{ \xi_{\mu}D_{\tau}v^{\mu}+2\xi^{\mu}\partial_{\mu}\ln(a_{2}r)\right\} _{0}=-\frac{2}{\sigma}\left[(\rho+p)(u^{\mu}\xi_{\mu})^{2}+p\right]_{0},
\end{equation}
and the energy conservation equation takes the form 
\begin{equation}
\left[(\rho+p)u_{\mu}v^{\mu}u_{\nu}\xi^{\nu}\right]_{0}=0.
\end{equation}

Let us first look at Eq. (\ref{junction1}). Outside the wall, we
have 
\begin{equation}
\xi^{\mu}\partial_{\mu}\ln(a_{2}r)=a_{1}r_{,\tau}\frac{\dot{a_{2}}}{a_{2}}+\frac{t_{,\tau}}{a_{1}}\left(\frac{a_{2}^{\prime}}{a_{2}}+\frac{1}{r}\right).
\end{equation}
Inside the wall, 
\begin{equation}
\xi^{\mu}\partial_{\mu}\ln(ar)=r_{,\tau}\dot{a}+\frac{t_{,\tau}}{ar}.
\end{equation}
Then Eq. (\ref{junction1}) yields 
\begin{equation}
a_{2}^{\prime}=-4\pi\sigma a^{2}\sqrt{1+a^{2}r_{,\tau}^{2}}.
\end{equation}

Similarly, manipulating Eq. (\ref{junction2}) gives 
\begin{equation}
a_{1}^{\prime}=\frac{4\pi\sigma\sqrt{1+a^{2}r_{,\tau}^{2}}}{r_{,\tau}^{2}}.
\end{equation}
These two equations can be rewritten as 
\begin{equation}
a_{1}^{\prime}=\frac{4\pi\sigma\sqrt{1-a^{2}\dot{r}^{2}}}{\dot{r}^{2}},
\end{equation}
\begin{equation}
a_{2}^{\prime}=-\frac{4\pi\sigma a^{2}}{\sqrt{1-a^{2}\dot{r}^{2}}}.
\end{equation}

Now we assume $\left[u_{\mu}\xi^{\mu}\right]_{0}=0.$ Since $u_{\mu}\xi^{\mu}$
is the 4-velocity of the fluid in the direction of the unit normal
vector, this condition means the fluid flows through the wall smoothly.
One then obtains 
\begin{equation}
\left\{ \xi_{\mu}D_{\tau}v^{\mu}+2\xi^{\mu}\partial_{\mu}\ln(a_{2}r)\right\} _{0}=0,
\end{equation}
which leads to the wall's equation of motion 
\begin{equation}
\frac{ar_{,\tau\tau}}{\sqrt{1+a^{2}r_{,\tau}^{2}}}+\frac{4a_{,\tau}r_{,\tau}}{\sqrt{1+a^{2}r_{,\tau}^{2}}}+\frac{2\sqrt{1+a^{2}r_{,\tau}^{2}}}{ar}=6\pi\sigma.
\end{equation}
This can also be written as 
\begin{equation}
\ddot{r}+(4-3a^{2}\dot{r}^{2})H\dot{r}+\frac{2}{a^{2}r}(1-a^{2}\dot{r}^{2})=6\pi\sigma\frac{(1-a^{2}\dot{r}^{2})^{\frac{3}{2}}}{a},\label{4-1}
\end{equation}
which is Eq. (\ref{4}).

Furthermore, we are able to determine the time when apparent horizon
arises. By Eq. (\ref{theta}), when $\Theta_{out}=0$, that is, $\dot{(a_{2}r)}+(a_{2}r)^{\prime}/a_{1}=0$,
one obtains 
\begin{equation}
\left(H+\frac{1}{ar}\right)\sqrt{\frac{1+a\dot{r}}{1-a\dot{r}}}=4\pi\sigma,
\end{equation}
which is Eq. (\ref{5}).

Lastly, the Misner-Sharp mass right outside the wall is 
\begin{eqnarray}
M & = & \frac{a_{2}r}{2}\left\{ 1-\frac{(a_{2}r)^{\prime}{}^{2}}{a_{1}^{2}}+\dot{(a_{2}r)}^{2}\right\} \\
 & = & \frac{1}{2}H^{2}(ar)^{3}+\frac{4\pi\sigma(ar)^{2}}{\sqrt{1-a^{2}\dot{r}^{2}}}+\frac{4\pi\sigma H\dot{r}}{r\sqrt{1-a^{2}\dot{r}^{2}}}(ar)^{4}-8\pi^{2}\sigma^{2}(ar)^{3}.\label{mass}
\end{eqnarray}
These four terms are the volume energy, surface energy, surface-volume
binding energy and surface-surface binding energy, respectively. When
$r$ satisfies Eq. (\ref{5}), $M=ar/2$.

If the spacetime inside the wall is Minkowski, $a=1$ and $H=0.$
As a critical solution of Eq. \eqref{4-1}, $r\to\text{const.}$,
which gives $r=\frac{1}{3\pi\sigma}.$ Then Eq. (\ref{mass}) gives
$M_{cr}=\frac{4}{27\pi\sigma},$ which is exactly the critical mass
obtained in Ref. \cite{Blau}.

\end{document}